\documentclass[showpacs,showkeys,prc,twocolumn]{revtex4}
\usepackage{amssymb}
\usepackage{amsmath}
\usepackage[dvipdfm]{graphicx}
\usepackage{dcolumn}
\usepackage{bm}
\usepackage{amsfonts}

\begin{document}

\title[Gamov]{Continuum corrections to the level density and its dependence
on excitation energy, \textit{n-p} asymmetry, and deformation.}
\author{R. J. Charity}
\author{L. G. Sobotka}
\affiliation{Department of Chemistry, Washington University, St. Louis, Missouri 63130.}

\begin{abstract}
In the independent-particle model, the nuclear level density is determined
from the neutron and proton single-particle level densities. The
single-particle level density for the positive-energy continuum levels is
important at high excitation energies for stable nuclei and at all
excitation energies for nuclei near the drip lines. This single-particle
level density is subdivided into compound-nucleus and gas components. Two
methods were considered for this subdivision. First in the subtraction
method, the single-particle level density is determined from the scattering
phase shifts. In the Gamov method, only the narrow Gamov states or
resonances are included. The level densities calculated with these two
methods are similar, both can be approximated by the backshifted Fermi-gas
expression with level-density parameters that are dependent on $A$, but with
very little dependence on the neutron or proton richness of the nucleus.
However, a small decrease in the level-density parameter was predicted for
some nuclei very close to the drip lines. The largest difference between the
calculations using the two methods was the deformation dependence on the
level density. The Gamov method predicts a very strong peaking of the level
density at sphericity for high excitation energies. This leads to a
suppression of deformed configurations and, consequently, the fission rate
predicted by the statistical model is reduced in the Gamov method.
\end{abstract}

\maketitle

\section{INTRODUCTION}

\label{sec:intro}The nuclear level density $\rho $ is an essential
ingredient in calculating the statistical decay of a compound nucleus (CN)
by particle evaporation, gamma-ray emission, or fission. The statistical
model has widespread use in nuclear physics and applied research. All told
in these areas, knowledge of the level density is needed at low and high
excitation energies, with small and large compound-nucleus spins, and for
the full range of $Z$\ and $N$ from around the $\beta $ line of stability
out towards the drip lines. For example, the cross sections for neutron
capture on nuclei close to the neutron and proton drip lines are of
interested in r and rp nucleosynthesis calculations. If the excitation
energy in these reactions is sufficient, the statistical model is used to
determined the \textit{n} and $\gamma $ decay rates of the fused system. In
such applications, level densities are required for nuclei with extreme 
\textit{n}-\textit{p} asymmeties. For many of these nuclei, it will not be
possible to measure the level density even with proposed radioactive beam
facilities. Clearly a good understanding of the dependence of the level
density on the \textit{n}-\textit{p} asymmetry is required to extrapolate to
these systems. Even for less exotic compound nuclei closer to the $\beta $
line of stability, an asymmetry dependence can have important consequences
on the \textit{n-p} asymmetry of the evaporation residues\cite{Charity03}.
For fission decay, the deformation dependence of the level density is also
needed. Therefore it is important to know the excitation-energy, asymmetry,
and deformation dependencies of the level density over most regions of the
chart of nuclides.

A complete understanding of the nuclear level density requires consideration
of the many-body nature of the nucleus. However, the independent-particle
model provides a useful reference to start with. It also permits a rapid
survey of level-density dependencies over many regions of the chart of
nuclides and gives insight into how different nuclear-structure effects
modify the level density. Many-body effects such as the effective-nucleon
masses and collective enhancement due to rotational and vibrational
collective modes can be incorporated in a phenomenological way onto the
independent-particle model\cite{Ignatyuk79,Ma83}. In this paper, the
excitation-energy, \textit{n}-\textit{p} asymmetry, and deformation
dependencies of the level density are investigated within the framework of
the independent-particle model. Specifically, the role of the continuum of
positive-energy single-particle states is studied. For systems around the $%
\beta $ line of stability these states are populated significantly only at
large excitation energies. However for systems closer to the drip lines,
where either the neutron or proton separation energy is small, these states
can be populated significantly even at low excitation energies. It has been
suggested that the contributions from these continuum states may lead to a 
\textit{n-p} asymmetry dependence of level density\cite{Al-Quraishi01}. The
manner in which these states influence the deformation and excitation-energy
dependencies of the level density will also be investigated.

Before further discussion of the level density, it is useful to first
consider the largest excitation energies for which it is meaningful to apply
the statistical model. The CN is a system of nucleons which is equilibrated
in its single-particle degrees of freedom and thus has a long lifetime
compared to the timescale of single-particle motion. As such,
compound-nucleus decay is a rare process, i.e., the typically energy
fluctuation of a nucleon, which is of order of the temperature $T$, does not
lead to the emission of that particle. Thus, the regime of applicability is $%
T<E_{cost}^{\min }$ where $E_{cost}^{\min }$ is the minimum of $E_{cost}^{n}$
and $E_{cost}^{p}$, the energetic costs of emitting a neutron or a proton,
respectively. For neutrons, the cost is just the neutron separation energy $%
E_{cost}^{n}$=$E_{sep}^{n}$, while for protons the cost also includes the
Coulomb barrier $E_{cost}^{p}$=$E_{sep}^{p}+V_{coul}$.

The decay width for protons or neutrons is roughly\cite{Siemens87}%
\begin{equation}
\Gamma =\frac{t^{2}}{\pi \epsilon _{0}}\exp \left( -E_{cost}/t\right)
\end{equation}%
where $\epsilon _{0}$=$\hbar ^{2}/2mR^{2}$, $1/t=d\rho /dE^{\ast }$ is the
nuclear temperature ($t\approx T$), $m$ is the nucleon mass, and $R$ is the
nuclear radius. As $t$ approaches $E_{cost}^{\min }$, the decay width for
either proton or neutron evaporation becomes quite large. For the
statistical model to be applicable, the total decay width must be small
compared to the spreading width which determines the time scale for the CN
to equilibrate.

The order of this paper is as follows. A brief review of the level density
in the independent-particle model is given in Sec.~\ref{sec:independent}.
Subsequently, two methods to include the positive-energy states are
considered in Sec.~\ref{sec:contimuum}. Following this in Sec.~\ref{sec::ccc}%
, the details of the coupled-channels calculation of the single-particle
level densities are given. The determination of the deformation energy is
discussed in Sec.~\ref{sec:deformation} and calculated level densities are
presented in Sec.~\ref{sec:calculations}. Finally in Secs.~\ref%
{sec:discussion} and \ref{sec:conclusion}, a discussion of the results and
the conclusions of this work are made.

\section{LEVEL DENSITY IN THE INDEPENDENT-PARTICLE MODEL}

\label{sec:independent} The independent-particle model starts with sets of
single-particle levels for both neutrons and protons. The determination of
the nuclear level density is essentially a combinatorial problem, i.e., how
many ways can these single-particle levels be occupied to give the desired
total excitation energy. However, the enumeration of all the single-particle
configurations can be avoided. Instead, the Laplace transform $Z(\alpha
,\beta )$ of the level density is more easily calculated when the Lagrange
multipliers $\alpha $ and $\beta $ are introduced to constraint the total
number of particles and the total energy. The inverse transform can be
obtained from the saddle-point approximation to give a formula for the level
density which is continuous in excitation energy $E^{\ast }$. For simplicity
at this point, consider only one type of nucleon with single-particle levels 
$\varepsilon _{i}$, the level density is then\cite{Bohr75} 
\begin{equation}
\rho (E^{\ast })=\frac{\exp S}{2\pi \sqrt{D}}  \label{eq:general_FG}
\end{equation}%
where $S=\beta E-\alpha A+\ln Z(\alpha ,\beta )$. The values of the Lagrange
multipliers are determined by the saddle-point condition $\partial
S/\partial \beta $=$\partial S/\partial \alpha $=0. Now the average
occupancy of a single-particle level is given by $f_{i}=1/\left[ 1+\exp
\left( \beta \varepsilon _{i}-\alpha \right) \right] $. Thus the
saddle-point condition can be expressed in terms of the conservation of
nucleon number $A$\ and total energy $E=E_{gs}+E^{\ast }$ (ground-state +
excitation energy) by 
\begin{eqnarray}
A &=&\frac{\partial \ln Z}{\partial \alpha }=\sum_{i}f_{i},  \label{eq:A} \\
E &=&E_{gs}+E^{\ast }=-\frac{\partial \ln Z}{\partial \beta }%
=\sum_{i}\varepsilon _{i\,}f_{i}.  \label{eq:E}
\end{eqnarray}%
At the saddle point, the quantities $Z$, $D$, and $S$ are now%
\begin{eqnarray}
\ln Z &=&\sum_{i}\ln \left[ 1+\exp \left( \alpha -\beta \varepsilon
_{i}\right) \right] , \\
D &=&\left\vert 
\begin{array}{cc}
\frac{\partial ^{2}\ln Z}{\partial \alpha ^{2}} & \frac{\partial ^{2}\ln Z}{%
\partial \alpha \partial \beta } \\ 
\frac{\partial ^{2}\ln Z}{\partial \beta \partial \alpha } & \frac{\partial
^{2}\ln Z}{\partial \beta ^{2}}%
\end{array}%
\right\vert ,
\end{eqnarray}%
and%
\begin{equation}
S=\sum_{i}s_{i}.  \label{eq:S}
\end{equation}%
Here%
\begin{equation}
s_{i}=-f_{i}\ln f_{i}-(1-f_{i})\ln \left( 1-f_{i}\right) .
\label{eq:littles}
\end{equation}

Although this formula can be derived without recourse to statistical
mechanics, Bethe\cite{Bethe36} realized there is a close analogy to the
problem of a Fermi gas in contact with a heat bath of temperature $T$=1/$%
\beta $ and with chemical potential $\mu $=$\alpha /\beta $. In this
analogy, $Z$ is the grand partition function, $S$\ is the entropy and thus 1/%
$T=dS/dE^{\ast }$.

If the single-particle level-density $g(\varepsilon )=\sum_{i}\delta
(\varepsilon -\varepsilon _{i})$ is constant (at least in the vicinity of $%
\varepsilon $=$\mu $), then Eq.~\ref{eq:general_FG} can be reduced to the
well known Bethe or Fermi-gas expression\cite{Bethe36,LeCouter59,Bohr75}:%
\begin{eqnarray}
\rho (E^{\ast }) &=&\frac{\exp S}{\sqrt{48}E^{\ast }}, \\
S &=&2\sqrt{aE^{\ast }}=aT^{2},
\end{eqnarray}%
where $a$=$\frac{\pi ^{2}}{6}\,g(\mu )$ is the level-density parameter. For
a two-component Fermi gas, the level density parameter will have
contributions from each component $a$=$\frac{\pi ^{2}}{6}\left[ g_{n}(\mu
_{n})+g_{p}(\mu _{p})\right] $. Experimentally, level-density parameters
exhibit strong shell corrections at low excitation energies. However apart
from this, the average value of the level-density parameter is often assumed
to depend only linearly on $A$ with no dependence on the \textit{n}-\textit{p%
} asymmetry.

\begin{figure}[tbp]
\includegraphics*[scale=.4]{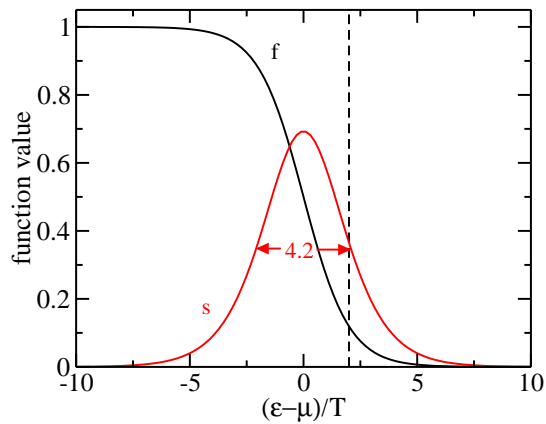}
\caption{(Color online) The evolution with ($\protect\varepsilon -\protect%
\mu )/T$ of the functions $f$ and $s$ (Eq.~\protect\ref{eq:littles}) used to
determine the total energy and entropy. The vertical dashed line indicates
the location of the $\protect\varepsilon $=0, when $T$=$E_{sep}$/2.}
\label{fig:fermi}
\end{figure}

To gauge the temperatures for which this formula should be applied, the
functions $f$ (Fermi function) and $s,$ which are needed to determined the
total energy and entropy (Eqs.~\ref{eq:E} and \ref{eq:S}), are plotted in
Fig.~\ref{fig:fermi} verses $\beta \varepsilon -\alpha $=$\left( \varepsilon
-\mu \right) /T$. The Fermi function $f$, giving the average level
occupancy, changes in value from 90\% to 10\% over an interval $\Delta
\varepsilon $=4.4$T$ centered around $\mu $. The function $s$ is
Gaussian-like with a full width half maximum (FWHM) of 4.2$T$, however the
tails of the function falls off much slower than a Gaussian. The Fermi-gas
formula thus assumes the single-particle level density $g$ is constant at
least over an interval $\pm $2$T$ around $\mu $. However because $s$ falls
off so slowly, the contribution to the entropy from levels at smaller and
larger energies are not insignificant. Therefore at large temperatures, how
useful is the Fermi-gas formula when $g$ is not constant? At low
temperatures by expanding Eqs.~\ref{eq:A}, \ref{eq:E}, and \ref{eq:S} as
functions of $T$, the entropy with its lowest-order correction becomes $S=%
\sqrt{a^{\prime }E^{\ast }}$ where 
\begin{equation}
a^{\prime }=a\left[ 1+\frac{7g(\mu _{0})g^{\prime \prime }(\mu
_{0})-5g^{\prime }(\mu _{0})^{2}}{5g(\mu _{0})^{3}}E^{\ast }\right] ,
\label{eq:fermiplus}
\end{equation}%
$\mu _{0}$ is the chemical potential at $T$=0, and $g^{\prime }$ and $%
g^{\prime \prime }$ are the first and second derivatives of $g$. Thus the
level-density parameter can be replaced by an effective level-density
parameter $a^{\prime }$ which is excitation-energy dependent. Higher-order
corrections will be needed at larger temperatures.

What about the role of positive energy states? Consider a system where $T$ =$%
E_{sep}/2$. In Fig.~\ref{fig:fermi} for this case, $\epsilon $=0 would
correspond to the vertical dashed line and clearly, because $s$ decreases so
slowly with $\varepsilon $, the positive-energy states (those beyond the
dash line) are important for the entropy.

\section{CONTINUUM SINGLE-PARTICLE LEVEL DENSITY}

\label{sec:contimuum}

In order to understand the role of positive-energy states in calculating the
level density, we need a prescription for deciding which of these
positive-energy levels belong to the CN. If there is no bounding volume
enclosing the nucleus, the single-particle level density of these states is
infinite. However not all of these states are considered to be associated
with the CN. To better understand the selection of positive-energy states
belonging to the CN, one can consider the problem of nucleon evaporation.
This is often dealt with by placing the CN in a box whose volume is large
compared to the nuclear volume. Call this state \textit{i}. The box volume $%
V $ can eventually be expanded and allowed to approach infinity. If the CN
decays by the emission of a nucleon with kinetic energy $\varepsilon $ to
state \textit{f}, then as we have a bounding box, the nucleon cannot escape
and will eventually be reabsorbed by the daughter nucleus leading us back to
state \textit{i}. Therefore by the general principle of detailed balance,
the transition probability $P_{if}$ from state \textit{i} to \textit{f} is
related to that of the inverse process by 
\begin{equation}
\rho _{i}P_{if}=\rho _{f}P_{fi}
\end{equation}%
where $\rho _{i}$ and $\rho _{f}$ are the density of states for \textit{i}
and \textit{f,} respectively. For state \textit{i}, the density of states is
just the level density of the compound nucleus $\rho _{i}=\rho _{CN}(E^{\ast
})$. While for state \textit{f}, both the level density of the daughter
nucleus $\rho _{d}$ and the phase space $g_{ev}$ of the evaporated particle
contribute, .i.e., 
\begin{equation}
\rho _{f} =\rho _{d}(E^{\ast }-E_{sep}-\varepsilon )g_{ev}\left( \varepsilon
\right) d\varepsilon .
\end{equation}
Now 
\begin{eqnarray}
g_{ev}(\varepsilon ) &=&\left( 2s+1\right) \frac{4\pi p^{2}}{h^{3}}V\frac{dp%
}{d\varepsilon }  \notag \\
&=&\left( 2s+1\right) \frac{\left( 2m\right) ^{3/2}V\sqrt{\varepsilon }}{%
4\pi ^{2}\hbar ^{3}}  \label{eq:gev}
\end{eqnarray}%
where $p$, $m$ and $s$ are the evaporated nucleon's momentum, mass and spin,
respectively. Here $g_{ev}$ is determined for an empty box in the
semiclassical limit. This should be appropriate as the box volume is large.
Because $P_{\mathit{fi}}=v\,\sigma _{inv}\left( \varepsilon \right) /V$,
then 
\begin{eqnarray}
P_{\mathit{if}} &=&\frac{\Gamma \left( \varepsilon \right) d\varepsilon }{%
\hbar }  \notag \\
&=&\frac{(2s+1)m}{\left( \pi \hbar \right) ^{2}}\varepsilon \sigma
_{inv}\left( \varepsilon \right) \frac{\rho _{d}\left( E^{\ast
}-E_{sep}-\varepsilon \right) }{\rho _{CN}\left( E^{\ast }\right) }%
d\varepsilon
\end{eqnarray}%
where $v$ is the nucleon velocity and $\sigma _{inv}$ is the inverse or
absorption cross section.

In this derivation of the Weisskopf evaporation formula, the single-particle
level density contributes to both $\rho _{CN}$ and the phase space of the
evaporated nucleon $g_{ev}$. Thus for a given nuclear mean-field potential
surrounded by a bounding box, the total single-particle level density will
be subdivided; $g_{tot}\left( \varepsilon \right) =g_{CN}\left( \varepsilon
\right) +g_{gas}\left( \varepsilon \right) $ where $g_{CN}$ is the
single-particle level density used to calculate the compound-nucleus level
density and the remaining level density $g_{gas}$ is associated with a gas
of evaporated particles. Thus $g_{gas}\sim g_{ev}$, where $g_{ev}$ is the
single-particle level density for the empty bounding box, i.e., without the
nuclear mean-field potential (Eq.~\ref{eq:gev}). As the box volume is chosen
to be much larger than the nuclear volume, then we also find $g_{tot}\sim
g_{ev}$, though of course $g_{tot}\neq g_{gas}$. With such a subdivision of $%
g_{tot}$, the nucleon number in the box can be subdivided, i.e., 
\begin{eqnarray}
A_{tot} &=&\int g_{tot}\left( \varepsilon \right) d\varepsilon =\int
g_{CN}\left( \varepsilon \right) d\varepsilon +\int g_{gas}\left(
\varepsilon \right) d\varepsilon  \notag \\
&=&A_{CN}+A_{gas}.
\end{eqnarray}%
Similarly $E_{tot}=E_{CN}+E_{gas}$ and $S_{tot}=S_{CN}+S_{gas}$. For a given
temperature, the chemical potential $\mu $ is constrained so that $A_{CN}$
is the constant value appropriate for the CN. Thus $A_{gas}$ and $A_{tot}$
will be temperature dependent and hence $1$/$T=dS_{CN}/dE_{CN}$ while $%
dS_{tot}/dE_{tot}\neq $ $1$/$T\neq dS_{gas}/dE_{gas}.$

The quantity $g_{CN}$ should contain all the negative-energy bound states
located in the well of the nuclear- plus-Coulomb potential. For positive
energies, $g_{CN}$ should be independent of the bounding volume. However
apart from these constraints, there is no well justified subdivision of $%
g_{tot}$ into its two components in the independent-particle model. Two
methods have been utilized to calculate $g_{CN}$.

\subsection{Subtraction Method}

In 1978 Fowler, Engelbrecht, and Woosley\cite{Fowler78} proposed that $%
g_{gas}\equiv g_{ev}$ for neutrons and thus $g_{CN}$ could be obtained from
subtraction, i.e., $g_{CN}=g_{tot}-g_{ev}$. We will call this the
subtraction method for determining $g_{CN}$ and it has been used by many
other investigators. It is rather easy to show that\cite%
{Beth37,Dean85,Shlomo97,Newton02}%
\begin{eqnarray}
g_{CN}^{sub}\left( \varepsilon \right) &=&\sum_{l,j}g_{\ell ,j}\left(
\varepsilon \right) ,  \label{eq:sub} \\
g_{\ell ,j}\left( \varepsilon \right) &=&(2j+1)\sum_{i}\delta \left(
\varepsilon -\varepsilon _{i}^{\ell ,j}\right)  \notag \\
&&+\frac{1}{\pi }(2j+1)\frac{d\delta _{\ell ,j}}{d\varepsilon }
\end{eqnarray}%
where $\delta _{\ell ,j}\left( \varepsilon \right) $ is the phase shift
associated with the scattering state of energy $\varepsilon $, orbital
angular momentum $\ell $, and total angular momentum $j$. The bound-state
energies are $\varepsilon _{i}^{\ell ,j}$. From Levinson's theorem\cite%
{Dean85}, 
\begin{equation}
\int_{-\infty }^{\infty }g_{\ell ,j}\left( \varepsilon \right) d\varepsilon
=0.
\end{equation}%
It is clear that for $\varepsilon >0$, $g_{\ell ,j}$ must have a net
negative contribution to balance out the positive contributions from the
bound states. However, for the large $\ell $ waves, this negative
contribution occurs at very large $\varepsilon $ values which are not
populated in the CN\cite{Shlomo97}. Thus the negative contributions are only
important for the lowest $\ell $-waves.

Near a resonance 
\begin{equation}
\frac{d\delta _{\ell ,j}}{d\varepsilon }=\frac{\Gamma ^{R}/2}{\left(
\varepsilon -\varepsilon ^{R}\right) ^{2}+\left( \Gamma ^{R}/2\right) ^{2}}
\end{equation}%
where $\varepsilon ^{R}$ is the resonance energy and $\Gamma ^{R}$ is its
width. In the limit as $\Gamma ^{R}\rightarrow 0$, $d\delta _{\ell
,j}/d\varepsilon \rightarrow \pi \,\delta \left( \varepsilon -\varepsilon
_{R}\right) $ and the resonance becomes equivalent to a bound state.
Therefore at positive energies, $g_{CN}^{sub}$ consists of series of
resonance peaks which, for low $\ell $-waves, sit on a negative background.

For protons, the single-particle level density is calculated from the
nuclear phase shift, not the total phase shift. Hence, the subtracted level
density is not actually $g_{ev}$, the contribution from the bounding volume
without any mean-field potential, as used for neutrons. In this case, the
subtracted level density is that from the bounding volume containing a
point-source Coulomb potential. For deformed systems, the $\varepsilon >$0
contribution can be generalized as\cite{Tsang75,Osborn76,Osborn77} 
\begin{equation}
g_{CN}^{sub}(\varepsilon )=\frac{1}{2\pi i}\mathrm{Tr}\left( \bm{S}%
^{-1}(\varepsilon )\frac{d}{d\varepsilon }\bm{S}(\varepsilon )\right)
\label{eq:sub_Smatrix}
\end{equation}%
where $\bm{S}(\varepsilon )$ is the S-matrix for scattering at energy $%
\varepsilon $.

\begin{figure}[tbp]
\includegraphics*[scale=.35]{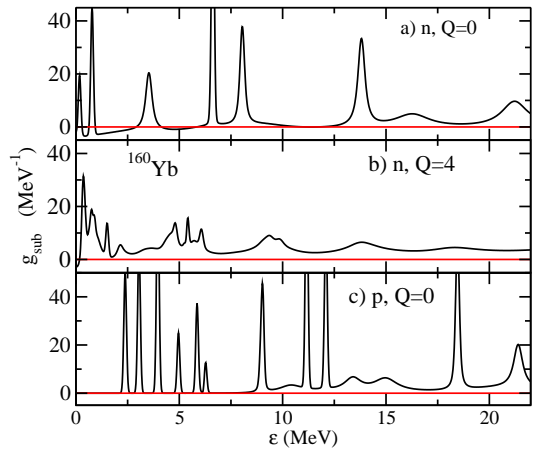}
\caption{(Color online) The dependence of the compound-nucleus
single-particle level density $g^{sub}_{CN}$ on the nucleon energy $\protect%
\varepsilon $. The displayed results have been convoluted by a Gaussian
resolution of FHWM=150~keV. Results are shown for a) neutron in a
spherically symmetric potential (Q=0), b) neutrons in a deformed potential
(Q=4), and c) protons in a spherically symmetric potential. }
\label{fig:gsub}
\end{figure}

Examples of $g_{CN}^{sub}\left( \varepsilon \right) $ calculated for $^{160}$%
Yb are shown in Fig.~\ref{fig:gsub}. The negative background is clearly
observable in Fig.~\ref{fig:gsub}a for neutrons in a spherically-symmetric
potential. Here $g_{CN}^{sub}\left( \varepsilon \right) $ is negative
between the resonance peaks for $\varepsilon <$6~MeV. In deforming the
potential, degenerate resonances are split and this often leads to a filling
up of the negative background so that $g_{CN}^{sub}$ seldom drops below
zero. For the deformed example in Fig.~\ref{fig:gsub}b, $g_{CN}^{sub}$ is
only negative at $\varepsilon \sim $0. For protons (Fig.~\ref{fig:gsub}c),
only very narrow resonances are observed well below the Coulomb barrier ($%
\varepsilon \sim $9~MeV). 

\subsection{Gamov Method}

By placing the CN in a bounding box, we have produced an equilibrium model.
Now within this equilibrium model, any arbitrary subdivision of $g_{tot}$
into $g_{CN}$ and $g_{gas}$ can be considered. Whatever subdivision is made,
the inverse cross section $\sigma _{inv}$ must be chosen to describe the
absorption of nucleons from the \textquotedblleft gas\textquotedblright\
phase space into the \textquotedblleft compound-nucleus\textquotedblright\
phase space. However, in order for the equilibrium evaporation rate to be
equated to that of an isolated CN which is a nonequilibrium problem, $g_{CN}$
should be chosen such that $f(\varepsilon )g_{CN}\left( \varepsilon \right) $
also describes the nucleon energy-density when there is no gas present. As $%
g_{CN}^{sub}$ can be negative for some energies, the number of nucleons
ascribed to the CN in these single-particle levels is also negative. In the
equilibrium model, this does not pose a problem as we have an accompanying
gas in the box. The total number of nucleons in any energy range is always
positive, i.e., $g_{tot}>$0. It is just our prescription of dividing up $%
g_{tot}$ into $g_{CN}$ and $g_{gas}$ that gives rise to this problem.
However in the absence of the gas, it is not clear what physical
significance a negative value of $g_{CN}$ has.

Another criticism of the subtraction method is that it includes both narrow
(long-lived) and wide (short-lived) resonances. It has been suggested that
only resonances of lifetime longer than, or comparable to, the
compound-nucleus lifetime should be included\cite{Weidenmuller64,Mustafa92}
as the occupancy of the shorter-lived levels will not be maintained before
the CN\ decays.

Consider the analytical continuation of the S-matrix into the complex-energy
plane. Poles of the S-matrix at $\varepsilon $=$\varepsilon ^{R}-i\Gamma
^{R}/2$ correspond to exponentially decaying solutions to the Schr\"{o}%
dinger equation (if $\Gamma ^{R}>$0)\cite{Bohm79,Newton02}. These are also
known as Gamov\cite{Gamov28} or Siegert states. For those poles close to the
real axis on the unphysical sheet, these states have a close association
with resonances. Weidenm\"{u}ller\cite{Weidenmuller64} suggested the
compound-nucleus single-particle level density for $\varepsilon >$0 should
be the density of sharp resonances or Gamov states. One should therefore
introduce a cutoff or maximum width $\Gamma _{0}$ of the Gamov states that
contribute to $g_{CN}$. Now, if the single-particle potential is modified,
for example by deformation, the location of the poles will move in the
complex-energy plane. Some will become narrower and some wider and a number
of these will cross the cutoff region causing discontinuities in the level
density as it evolves with deformation. In order to avoid such
discontinuities, a smooth exponential cutoff of the Gamov states was
implemented;%
\begin{equation}
g_{CN}^{\Gamma }(\varepsilon )=\sum_{i}\delta \left( \varepsilon
-\varepsilon _{i}^{R}\right) \exp \left( \frac{-\Gamma _{i}^{R}}{\Gamma _{0}}%
\right) .  \label{eq:g_gamov}
\end{equation}%
The summation includes all poles associated with bound states ($\Gamma ^{R}$%
=0) and Gamov states ($\Gamma ^{R}>$0). With this definition, $%
g_{CN}^{\Gamma }$ is always positive and thus avoids the ambiguities
associated with negative values as in $g_{CN}^{sub}$. In the following
sections, both definitions of $g_{CN}$ will be used to see how they affect
the excitation-energy, deformation, and asymmetry dependencies of the level
density.

\section{COUPLED-CHANNELS CALCULATION OF SINGLE-PARTICLE LEVELS}

\label{sec::ccc}

\subsection{Theory}

\label{sec::cctheory}

In order to calculate the single-particle level densities, the Schr\"{o}%
dinger equation must be solved to determine the bound, Gamov, and scattering
states. Protons and neutrons are assumed to move in an axially-symmetric
mean-field potential which is the sum of the nuclear, Coulomb, and spin-obit
components, i.e.,%
\begin{equation}
V(\bm{r})=V_{N}(\bm{r})+V_{C}(\bm{r})+V_{so}(\bm{r}).
\end{equation}%
The nuclear potential is expressed in terms of the Fermi function $f(x)=%
\left[ 1+\exp (x)\right] ^{-1}$ as%
\begin{equation}
V_{N}(\bm{r})=-V_{N}^{(0)}f\left( \frac{r-R\left( Q,\theta \right) }{d\left(
\theta \right) }\right)
\end{equation}%
where $R\left( Q,\theta \right) $ defines a spheroidal surface with the same
volume as a sphere of radius $R^{(0)}$. The deformation is express in terms
of the relative quadrupole moment $Q$ related to the radii, $r_{\parallel }$
and $r_{\perp }$, perpendicular and parallel to the symmetry axis by \cite%
{Hasse88}%
\begin{equation}
Q=\frac{8\pi }{15}\frac{\left( r_{\parallel }^{2}-r_{\perp }^{2}\right) }{%
\left[ R^{(0)}\right] ^{2}}.
\end{equation}%
The quadrupole moment is positive for prolate shapes, negative for oblate,
and zero at sphericity. As a calibration point, $Q$=3.2 corresponds to a
\textquotedblleft superdeformed\textquotedblright\ prolate shape with the
length of the major and minor axes differing by a factor of 2.

The diffuseness of the nuclear potential is assumed to be constant
perpendicular to this surface, i.e.,%
\begin{equation}
d\left( \theta \right) =d^{(0)}\sqrt{1+\left( \frac{dR}{d\theta }\frac{1}{R}%
\right) ^{2}}.  \label{eq:diffuseness}
\end{equation}%
The deformed spin-orbit interaction can be expressed in terms of the
momentum $\bm{p}$ and spin $\bm{s}$ operators as\cite{Esbensen00}%
\begin{equation}
V_{so}\left( \bm{r}\right) =4V_{so}^{(0)}\left( \left[ \bm{\nabla}f(\frac{%
r-R_{so}\left( Q,\theta \right) }{d_{so}\left( \theta \right) })\right]
\times \bm{p}\right) \cdot \bm{s}
\end{equation}%
where $d_{so}$ is defined in terms of $R_{so}$ in an equivalent manner as in
Eq.~\ref{eq:diffuseness}. The Coulomb potential is approximated as that from
a sharp-surfaced spheroid of equivalent spherical radius $R_{C}$ using the
analytical expressions of Refs.~\cite{Binney87,Chandrasekhar87}. The
parameters $V_{N}^{(0)}$, $V_{so}^{(0)}$, $R^{(0)}$, $R_{so}^{(0)}$, $R_{C}$%
, and $d^{(0)}$ are taken from the \textquotedblleft
universal\textquotedblright\ parametrization of Ref.~\cite{Dudek:82}.

The solutions to the Schr\"{o}dinger equation $H\Psi =\varepsilon \Psi $ are
obtained by expressing the wavefunction as sums of spherical waves specified
by $|\ell \frac{1}{2}jm\rangle $. Here $\ell $ is the orbital angular
momentum, $j$ is the total angular momentum and $m$ is its projection on the
symmetry axis. This latter quantity is conserved in the axially-symmetric
potential. Thus%
\begin{equation}
\Psi _{m}\left( \bm{r}\right) =\sum_{\ell ,j}\frac{u_{\ell jm}\left(
r\right) }{r}|\ell \frac{1}{2}jm\rangle
\end{equation}%
where $u_{\ell jm}\left( r\right) $ are the radial wavefunctions. After
projecting on the state $|\ell ^{\prime }\frac{1}{2}j^{\prime }m\rangle $,
the Schr\"{o}dinger equation can be written in terms of the coupled-channels
equation%
\begin{widetext}
\begin{equation}
\left[ \frac{d^{2}}{dr^{2}}+k^{2}-\frac{\ell \left( \ell +1\right) }{r^{2}}%
\right] u_{\ell jm}\left( r\right) +\sum_{\ell ^{\prime
},\,j^{\prime }}\left( W_{\ell j\ell ^{\prime }j^{\prime }}^{m}\left(
r\right) +D_{\ell j\ell ^{\prime }j^{\prime }}^{m}\left( r\right) \frac{d}{dr%
}\right) u_{\ell ^{\prime }j^{\prime }m}\left( r\right) =0.
\end{equation}%
\end{widetext}Here $k=\sqrt{2\mu \epsilon }/\hbar $ is the wave number and $%
\mu $ is the reduced mass. The matrices $\bm{W}$ and $\bm{D}$ are determined
from the matrix elements of the interaction taken between states specified
by $\ell ,j$ and $\ell ^{\prime },j^{\prime }$. The matrix $\bm{W}$ has
contributions from all three potentials (nuclear, Coulomb, and spin-orbit)
while $\bm{D}$ is determined only from the spin-orbit potential.

The boundary conditions at the origin are $u_{\ell jm}(0)$=0. If one
considers $N$ channels and chooses $N$ initial sets of the derivatives $%
du_{\ell jm}/dr(0)$ appropriately, then after integrating out from the
origin, one can obtain $N$ independent solutions to the coupled-channels
equation. Let these be represented by the $N$ columns of the $N\times N$
matrix $\bm{U^m}(r)$. In matrix form, the Schr\"{o}dinger equation is then 
\begin{equation}
\left[ \frac{d^{2}}{dr^{2}}+\bm{D^m}(r)\frac{d}{dr}+\bm{A^m}(r)\right] %
\bm{U^m}(r)=0  \label{eq:schrodinger}
\end{equation}%
where 
\begin{equation}
A_{\ell j\ell ^{\prime }j^{\prime }}^{m}=\left[ k^{2}-\frac{\ell \left( \ell
+1\right) }{r^{2}}\right] \delta _{\ell \ell ^{\prime }}\delta _{jj^{\prime
}}+W_{\ell j\ell ^{\prime }j^{\prime }}^{m}.
\end{equation}%
The equation is integrated out to a radius $r_{\mathrm{match}}$ where $V$$%
\rightarrow $0 for neutrons or, for protons, only a point-source Coulomb
term is present. At $r_{\mathrm{match}}$, the solutions are matched to
specific solutions of the Schr\"{o}dinger equations $p_{\ell }(r)$. For
bound states, the matching solution must vanish as $r$$\rightarrow $$\infty $
and thus $p_{\ell }(r)$=$\sqrt{2\left\vert k\right\vert r/\pi }\,K_{\ell +%
\frac{1}{2}}(\left\vert k\right\vert r)$ or $p_{\ell }(r)$=$W_{-\eta ,\ell +%
\frac{1}{2}}(2\left\vert k\right\vert r)$ for neutrons and protons,
respectively. Here $K_{\ell +\frac{1}{2}}$ are the modified Bessel functions
of the second kind and $W_{-\eta ,\ell +\frac{1}{2}}$ are the Whittaker
functions and $\eta =(Z-1)e^{2}/(\hbar ^{2}\left\vert k\right\vert )$. For
Gamov states, the matching functions are outgoing waves; $p_{\ell }(r)$=$kr%
\left[ j_{\ell }(kr)+i\,y_{\ell }(kr)\right] $ or $p_{\ell }(r)$=$F_{\ell
}(\eta ,kr)$ $-i\,G_{\ell }(\eta ,kr)$ for neutrons and protons
respectively. Here $j_{\ell }$ and $y_{\ell }$ are the regular and irregular
spherical Bessel functions and $F_{\ell }$ and $G_{\ell }$ are regular and
irregular Coulomb wavefunctions. If the calculated solution is to represent
a bound or Gamov \ state, then one must be able match the logarithmic
derivatives of $u_{\ell jm}$ and $p_{l}$ at $r=r_{\mathrm{match}}$ for all
channels. Any linear combination of the column vectors of $\bm{U^m}$ can be
used to achieve this match and it is only possible when\cite{Johnson78} 
\begin{equation}
\left\vert \frac{d\bm{U^m}}{dr}\left( \bm{U^m}\right) ^{-1}-\frac{d\bm{P}}{dr%
}\left( \bm{P}\right) ^{-1}\right\vert _{r=r_{\mathrm{match}}}=0
\end{equation}%
where the matrix $\bm{P}$ is defined as 
\begin{equation}
P_{\ell j\ell ^{\prime }j^{\prime }}(r)=\delta _{\ell \ell ^{\prime }}\delta
_{jj^{\prime }}p_{\ell }(r).
\end{equation}%
The matrix $\bm{Y}=$ d$\bm{U^m}/dr\left( \bm{U^m}\right) ^{-1}$ is called
the log-derivative matrix and satisfies the following Ricatti equation%
\begin{equation}
\frac{d\bm{Y}}{dr}+\bm{A}+\bm{Y}^{2}+\bm{D}\bm{Y}=0.
\end{equation}%
Rather than solving the matrix Schr\"{o}dinger equation (Eq.~\ref%
{eq:schrodinger}), this equation can be solved directly using the techniques
of Refs.~\cite{Johnson73,manolopoulos86,Mrugala83}. In fact, it is
advantageous to solve the Ricatti equation instead of the Schr\"{o}dinger
equation as the latter suffers from numerical instabilities when integrating
over classically forbidden regions.

To obtain scattering solutions, the wavefunctions must be matched to a
combination of ingoing and outgoing waves at $r=r_{\mathrm{match}}$. The
scattering matrix can also be obtained directly from the log-derivative\cite%
{Johnson73}. Defining the matrix elements%
\begin{align}
J_{\ell j\ell ^{\prime }j^{\prime }}(r)& =\delta _{\ell \ell ^{\prime
}}\delta _{jj^{\prime }}\,kr\,j_{\ell }(kr)\;\text{for neutrons}  \notag \\
& =\delta _{\ell \ell ^{\prime }}\delta _{jj^{\prime }}\,F_{\ell }(kr)\text{%
\ \ \ \thinspace for protons} \\
N_{\ell j\ell ^{\prime }j^{\prime }}(r)& =\delta _{\ell \ell ^{\prime
}}\delta _{jj^{\prime }}\,kr\,y_{\ell }(kr)\;\text{for neutrons}  \notag \\
& =-\delta _{\ell \ell ^{\prime }}\delta _{jj^{\prime }}\,G_{\ell }(kr)\text{%
\thinspace\ for protons,}
\end{align}%
the $\bm{K}$ matrix is determined by%
\begin{eqnarray}
\bm{K} &=&-\left[ \bm{Y}(r_{\mathrm{match}})\bm{N}(r_{\mathrm{match}})-\frac{%
d}{dr}\bm{N}(r_{\mathrm{match}})\right] ^{-1}  \notag \\
&&\times \left[ \bm{Y}(r_{\mathrm{match}})\bm{J}(r_{\mathrm{match}})-\frac{d%
}{dr}\bm{J}(r_{\mathrm{match}})\right] .
\end{eqnarray}%
The $\bm{S}$ matrix is derived in terms of the identity matrix $\bm{I}$ as 
\begin{equation}
\bm{S}=\left( \bm{I}+i\,\bm{K}\right) ^{-1}\left( \bm{I}-i\,\bm{K}\right) .
\end{equation}

The calculation of $g_{CN}^{sub}$ from Eq.~\ref{eq:sub_Smatrix} can be
problematic near very narrow resonances. However, the level density
convoluted with a small dispersion is more easily determined. If $%
F(\varepsilon )\ $is the convolution function, the convoluted level density
is 
\begin{equation}
\widetilde{g_{c}}(\varepsilon )=\int_{0}^{\infty }g_{c}(\varepsilon ^{\prime
})F(\varepsilon -\varepsilon ^{\prime })d\varepsilon ^{\prime }.
\end{equation}%
Following Sandulescu \textit{et al}.\cite{Sandulescu96}, the integral along
the real axis can be replaced by a contour integral in the complex-energy
plane which avoids narrow resonances. The contour $C$ is chosen to follow
the real axis except near resonances with $\Gamma <$ 50 keV where it follows
a semi-circular path of radius 0.2 MeV around each resonance. From Cauchy's
theorem, the final level density is 
\begin{equation}
\widetilde{g_{c}}(\varepsilon )=\sum_{n}F\left( \varepsilon -\varepsilon
_{n}^{R}\right) +\int_{C}g_{c}(\varepsilon ^{\prime })F(\varepsilon
-\varepsilon ^{\prime })d\varepsilon ^{\prime }
\end{equation}%
where here $\varepsilon _{n}=\varepsilon _{n}^{R}-i\Gamma _{n}^{R}/2$ is the
complex energy of the $n$th avoided resonance. The convolution function $F$
was taken as Gaussian with FWHM=150 MeV. This small resolution does not have
any significant affect on the deduced level densities in this work.

\subsection{Results}

\label{sec:results}

\begin{figure}[tbp]
\includegraphics*[scale=.4]{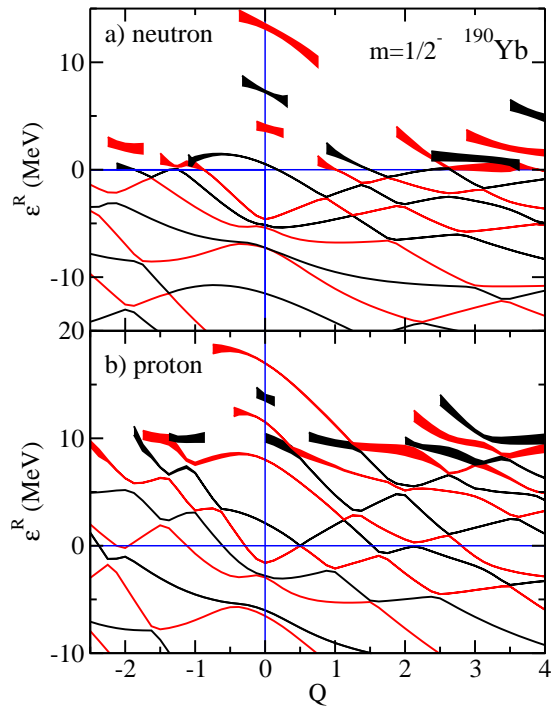}
\caption{(Color online) Evolution of the real energy $\protect\varepsilon %
^{R}$ for bound ($\protect\varepsilon ^{R}<$0, $\Gamma ^{R}$=0) and narrow
Gamov $\left( \protect\varepsilon ^{R}>\text{0, }\Gamma ^{R}<\text{1~MeV}%
\right) $ states as a function of deformation. The width of the curves
reflects the value of $\Gamma ^{R}$. Results are only shown for a) neutron
and b) proton $m^{\protect\pi }$=1/2$^{-}$ states in $^{190}$Yb.}
\label{fig:spaghetti}
\end{figure}

An example of the evolution of bound single-particle levels and the real
part of narrow Gamov states ($\Gamma <$0.5~MeV) with deformation is shown in
Fig.~\ref{fig:spaghetti} for $m^{\pi }$=$\frac{1}{2}^{-}$ neutrons in $%
^{190} $Yb. The results were obtained by including all channels with $\ell
\leq $20 in the coupled-channels calculations. The bound states and
resonances levels both move around with deformation, but levels of the same $%
m^{\pi }$ values avoid crossing each other. Bound levels that pass through $%
\varepsilon $=0 immediately become narrow resonances and vice versa. As the
resonance energy increases, the width of a resonance generally increases as
shown in Fig.~\ref{fig:spaghetti}.

The behavior that bound states turn into narrow resonances is quite general
except if there is no barrier (centrifugal or Coulomb)\cite{Newton02}. For $%
j $=$\frac{1}{2}^{+}$($\ell $=0) neutrons there is no barrier and bound
states passing through $\varepsilon $=0 turn into virtual states\cite%
{Newton02}. A virtual state is associated with a pole of the S matrix on the
real $\varepsilon $ axis at energy $\varepsilon _{v}=-e_{v}$ ($e_{v}>0$ and
small). Both bound and virtual states have real negative energies and purely
imaginary wavenumbers $k$. However for bound states, the imaginary part of $%
k $ is positive while it is negative for virtual states. In fact when there
is no barrier, both bound and virtual states with small energies have
important influences on the scattering at small positive values of $%
\varepsilon $. This has implications for the single-particle level density
calculated with the subtraction method. Consider a spherically symmetric
potential. The $\ell $=0 contribution to the single-particle level density
from a virtual state for small $\varepsilon $ values is\cite{Newton02} 
\begin{equation}
g_{0}^{sub}(\varepsilon )=%
\begin{cases}
\frac{1}{2\pi }\sqrt{\frac{e_{v}}{\varepsilon }}\frac{1}{\varepsilon +e_{v}}
& \text{if }\varepsilon >0, \\ 
0 & \text{if }\varepsilon <0.%
\end{cases}
\label{eq:virtual}
\end{equation}%
On the other hand for a bound state at $\varepsilon _{b}=-e_{b}$ ($e_{b}>0$
and small), the contribution is\cite{Newton02} 
\begin{equation}
g_{0}^{sub}(\varepsilon )=%
\begin{cases}
-\frac{1}{2\pi }\sqrt{\frac{e_{b}}{\varepsilon }}\frac{1}{\varepsilon +e_{b}}
& \text{if }\varepsilon >0 \\ 
=\delta (\varepsilon +e_{b}) & \text{if }\varepsilon <0.%
\end{cases}
\label{eq:bound}
\end{equation}%
As $\varepsilon \rightarrow +0$, $g_{0}^{sub}\rightarrow +\infty $ for the
virtual state and $-\infty $ for the bound state. In the limit that $%
e_{v}\rightarrow 0$ and $e_{b}\rightarrow 0,$ then in both cases $%
g_{0}^{sub}(\varepsilon )\rightarrow \delta (\varepsilon )/2$ and this
represents half a level. Thus in the deformation region over which a bound
level becomes a virtual state, $g_{CN}^{sub}$ evolves smoothly.

\begin{figure}[tbp]
\includegraphics*[scale=0.4]{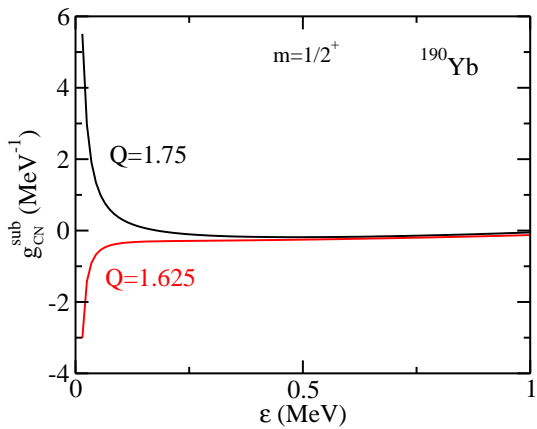}
\caption{(Color online) Variation with nucleon energy $\protect\varepsilon $
of the single-particle level density determined by the subtraction method
for $m^{\protect\pi }$=1/2$^{+}$ neutrons in $^{190}$Yb. Results are shown
for two neighboring deformations Q=1.75 and 1.625.}
\label{fig:virtual}
\end{figure}

Only the $m$=$\frac{1}{2}^{+}$states contain any $j$=$\frac{1}{2}^{+}$($\ell 
$=0) component in their wavefunction. For these states, the behavior as a
bound state passes through $\varepsilon $=0 is more complex. Sometimes they
become narrow resonances, sometimes they become virtual states, and other
times they progress in complicated manner to a wider resonance. As an
example, Figure~\ref{fig:virtual} shows the $m$=$\frac{1}{2}^{+}$
contribution to the single particle level-density $g_{1/2^{+}}^{sub}(%
\varepsilon )$ at small positive energies obtained for two neighboring
values of $Q$. For $Q$=1.625, $g_{1/2^{+}}^{sub}\rightarrow -\infty $ as $%
\varepsilon \rightarrow +0$ as in Eq.~\ref{eq:bound}. On the other hand, at $%
Q$=1.75, $g_{1/2^{+}}^{sub}\rightarrow +\infty $ as in Eq.~\ref{eq:virtual}.
In this case the behavior of $g_{1/2^{+}}^{sub}$is consistent with a bound
state at $Q$=1.625 passing through $\varepsilon $=0 and becoming a virtual
state at $Q$=1.75. 

For the Gamov method, $g_{1/2^{+}}^{\Gamma }(\varepsilon )$ does not evolve
smoothly when a bound state becomes a virtual state; bound states are always
counted as a full level while virtual states are not counted at all in Eq.~%
\ref{eq:g_gamov}. However, virtual states can cause long time delays in
scattering like narrow resonances\cite{Bohm79} and some thought should be
given to expanding the definition of $g_{CN}^{\Gamma }$ to include some
contribution from virtual states and so make the evolution with $Q$
smoother. In the present work this is not a significant issue as, in the
range of deformation investigated $\left( -2.5<Q<4\right) $, there is
typically only 2 small discontinuities.

\begin{figure}[tbp]
\includegraphics*[scale=0.4]{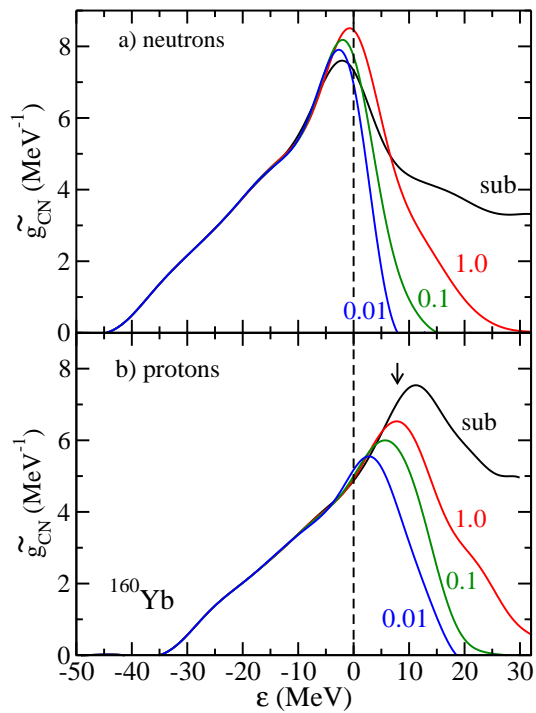}
\caption{(Color online) Smoothed single-particle level densities calculated
for a)\ neutrons and b) protons in $^{160}$Yb. The curves labeled
\textquotedblleft sub\textquotedblright\ were obtained from the subtraction
method. The other curves were obtained from the Gamov method with the
indicated values of $\Gamma _{0}$ in MeV. The Coulomb barrier for protons is
indicated by the arrow in b).}
\label{fig:smooth}
\end{figure}

To visualized the gross differences between the subtraction and Gamov
methods it is useful to smooth the single-particle level density $g_{CN}$.
Figure~\ref{fig:smooth} displays smoothed neutron and proton single-particle
level densities $\widetilde{g_{CN}}$ for $^{160}$Yb. The Strutinsky
smoothing discussed in Sec.~\ref{sec:deformation} was utilized. For
neutrons, $\widetilde{g_{CN}}$ peaks near $\varepsilon $=0. The peak is
lower in magnitude for the subtraction method due to the presence of the
negative background. As resonances at larger $\varepsilon ^{R}$ tend to have
larger widths, the Gamov method, which excludes these wide resonances, makes 
$\widetilde{g_{CN}}$ drop quickly to zero for $\varepsilon \gg $0. The
effect is more pronounced the smaller the cutoff width $\Gamma _{0}$.
Protons exhibit similar behavior except they peak closer to the Coulomb
barrier whose magnitude is indicated by the arrow in Fig.~\ref{fig:smooth}b. 
\begin{figure}[tbp]
\includegraphics*[scale=0.4]{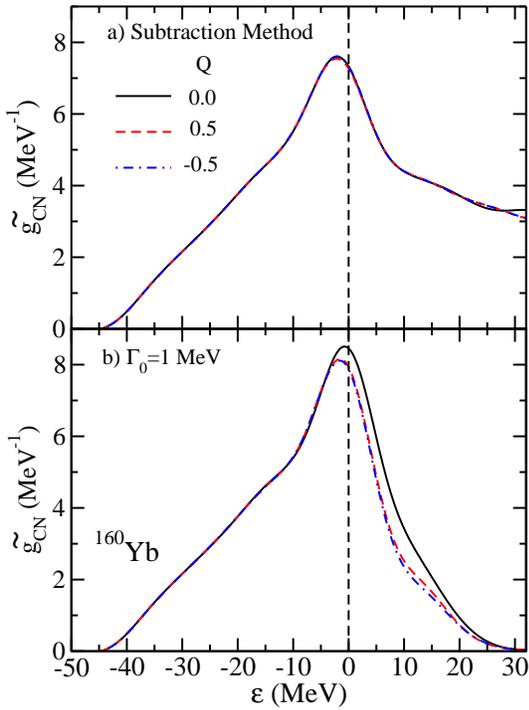}
\caption{(Color online) Smoothed single-particle level densities calculated
for\ neutrons in $^{160}$Yb at three defomations; $Q$=$-0.5,0,$ and 0.5. In
a)\ results obtained with the subtraction method are displayed while in b),
the Gamov method with $\Gamma _{0}$=1~MeV was used.}
\label{fig:smoothq}
\end{figure}

\begin{figure}
\includegraphics*[scale=0.4]{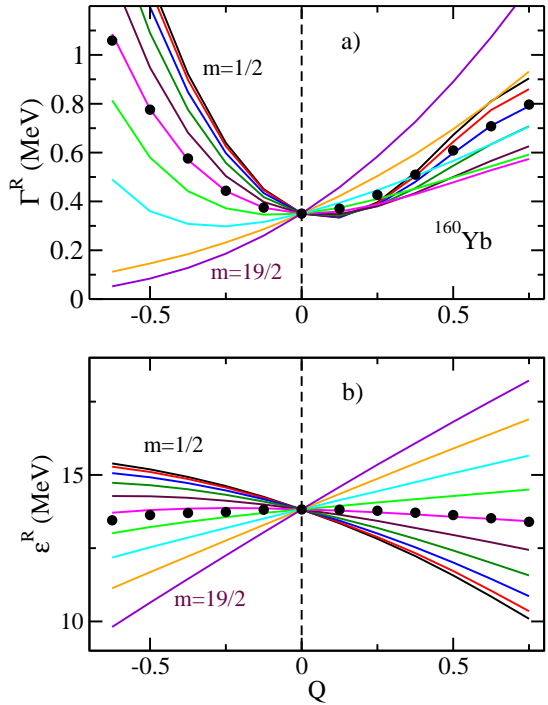}
\caption{(Color online) Curves show the variation with deformation $Q$ of
the energy $\protect\varepsilon ^{R}$ and width $\Gamma ^{R}$ of the Gamov
states associated with the $j$=19/2$^{-}$ neutrons in $^{160}$Yb. The solid
data points indicate the average energy and width for these states.}
\label{fig:width}
\end{figure}

There are two important results to highlight. First for protons, $%
g(\varepsilon )$ is almost independent of the method of calculation for
small positive energies well below the Coulomb barrier. These
positive-energy states have the most influence of the level density and thus
the continuum corrections for protons will generally be less important that
those for neutrons. Secondly for neutrons, the inclusion of the negative
background and the wide resonances in the subtraction method have opposite
effects and partially cancel. In the end, the level densities calculated
with both methods are found to be similar (see Sec.~\ref{sec:ex_dep}).

The subtraction method does not give a strong deformation dependence of $%
\widetilde{g_{CN}}$ near sphericity. As an example, the smoothed
single-particle level density for neutrons in $^{160}$Yb is plotted in Fig.~%
\ref{fig:smoothq}a for $Q$=$-0.5,0,0.5$. The curves for all three
deformations lie almost on top of each other. In contrast, the Gamov method
exhibits a strong dependence. The results, displayed in Fig.~\ref%
{fig:smoothq}b for $\Gamma _{0}$=1~MeV, show that $\widetilde{g_{CN}}$
decreases in magnitude for $\varepsilon >0$ as a deformation is imposed on
the compound nucleus (either prolate or oblate). To investigate this
behavior, let us concentrate on the splitting of Gamov states which are
degenerate at sphericity. For example in Fig.~\ref{fig:width}, the evolution
with deformation of the energy and width for a group of Gamov states
associated with $j$=19/2$^{-}$ neutrons in $^{160}$Yb is displayed. In Fig.~%
\ref{fig:width}b, the real energies $\epsilon ^{R}$ of the Gamov states
fanout with increasing deformation (both oblate and prolate). On average,
the mean value of $\epsilon ^{R}$ changes very little with deformation.
These mean values are indicated by the data points. In contrast, the widths $%
\Gamma ^{R}$ of Gamov states show a different behavior in Fig.~\ref%
{fig:width}b. Although a few of the states for oblate deformations show a
reduced width compared to sphericity, by in large the widths of most states
increase with deformation. The average widths, indicated by the data points,
have a minimum at sphericity. This behavior is typical of the splitting of
all degenerate Gamov states and therefore the average increase in these
widths with deformation reduces their contribution to $g_{CN}^{\Gamma }$
(Eq.~\ref{eq:g_gamov}). This explains the observed deformation dependence
displayed in Fig.~\ref{fig:smoothq}b. The strong dependence of $%
g_{CN}^{\Gamma }$ on deformation is reflected in the deformation dependence
of the level density (see Sec.~\ref{sec:deformation}).

\section{NUCLEAR LEVEL DENSITY WITH PAIRING}

\label{sec:LD_pair}

The simple discussion of the level density in Sec.~\ref{sec:independent} for
a single particle type and no interactions is extended in this section to
include both neutrons and protons and the pairing interaction. The grand
potential of a two-component Fermi gas is the sum of the proton and neutron
contributions, i.e.,%
\begin{equation}
\Omega \left( \alpha _{n},\alpha _{p},\beta \right) =\Omega _{n}\left(
\alpha _{n},\beta \right) +\Omega _{p}\left( \alpha _{p},\beta \right) .
\end{equation}%
The pairing interaction is considered in the BCS model\cite%
{Santo63,Moretto71,Jensen73}. In this model, the grand potential is related
to the grand partition function ($\Omega _{n}\left( \alpha _{n},\beta
\right) =-\ln Z_{n}/\beta )$ and for neutrons it is given by 
\begin{eqnarray}
\Omega _{n}\left( \alpha _{n},\beta \right) &=&\frac{\Delta _{n}^{2}}{G_{n}}%
+\int \frac{g_{n}(\varepsilon )}{2}\left[ \varepsilon -\mu _{n}-E\right]
d\varepsilon  \notag \\
&-&\frac{2}{\beta }\int \frac{g_{n}(\varepsilon )}{2}\ln \left[ 1+\exp
\left( -\beta E\right) \right] d\varepsilon
\end{eqnarray}%
where $\beta =1/T$, $T$ is the temperature, and $\mu _{n}=\alpha _{n}/\beta $
is the chemical potential. The quasiparticle energies are 
\begin{equation}
E=\sqrt{\left( \varepsilon -\mu _{n}\right) ^{2}+\Delta _{n}^{2}}.
\end{equation}%
The gap parameter $\Delta _{n}$ is determined from the gap equation 
\begin{equation}
\frac{2}{G_{n}}=\int \frac{g_{n}(\varepsilon )}{2}\frac{\tanh \left( \frac{E%
}{2T}\right) }{E}d\varepsilon  \label{eq:gap}
\end{equation}%
where $G_{n}$ is the pairing strength. The level density at an energy $%
E^{tot}$ can be obtained from the inverse Laplace transform of the grand
partition function 
\begin{widetext}
\begin{equation}
\rho \left( E^{tot},N,Z\right) =\frac{1}{\left( 2\pi i\right) ^{3}}%
\int_{-i\infty }^{+i\infty }\int_{-i\infty }^{+i\infty
}\int_{-i\infty }^{+i\infty }Z\exp \left( -\alpha _{n}N-\alpha
_{p}Z+\beta E\right) \,d\alpha _{n}\,d\alpha _{p}\,d\beta 
\end{equation}%
\end{widetext}which can be evaluated approximately by the saddle-point
method to give

\begin{equation}
\rho \left( E^{tot},N,Z\right) =\frac{\exp \left( S\right) }{\left( 2\pi
\right) ^{3/2}\sqrt{D}}.
\end{equation}%
Here the energy $\left( E^{tot}=E_{n}^{tot}+E_{p}^{tot}\right) $, entropy ($%
S=S_{n}+S_{p}$), and particle number\ are determined from the following
equations:%
\begin{align}
E_{n}^{tot}& =\int \varepsilon \frac{g_{n}(\epsilon )}{2}\left[ 1-\frac{%
\varepsilon -\mu _{n}}{E}\tanh \left( \frac{E}{2T}\right) \right]
d\varepsilon  \notag \\
& -\frac{\Delta _{n}^{2}}{G_{n}},  \label{eq:energy_tot} \\
S_{n}& =\int g_{n}(\varepsilon )\ln \left[ 1+\exp \left( -\frac{E}{T}\right) %
\right] d\varepsilon  \notag \\
& +\int g_{n}(\varepsilon )\frac{\frac{E}{T}}{1+\exp \left( \frac{E}{T}%
\right) }d\varepsilon , \\
N& =\int \frac{g_{n}(\varepsilon )}{2}\left[ 1-\frac{\varepsilon -\mu _{n}}{E%
}\tanh \left( \frac{E}{2T}\right) \right] \,d\varepsilon .
\end{align}%
The quantities $E_{p}^{tot}$, $S_{p}$, and $\Delta _{p}$ for protons are
obtained from similar expressions and the determinant $D$ is now 
\begin{equation}
D=\left\vert 
\begin{array}{ccc}
\frac{\partial ^{2}\ln Z}{\partial \alpha _{n}^{2}} & \frac{\partial ^{2}\ln
Z}{\partial \alpha _{n}\partial \alpha _{p}} & \frac{\partial ^{2}\ln Z}{%
\partial \alpha _{n}\partial \beta } \\ 
\frac{\partial ^{2}\ln Z}{\partial \alpha _{n}\partial \alpha _{p}} & \frac{%
\partial ^{2}\ln Z}{\partial \alpha _{p}^{2}} & \frac{\partial ^{2}\ln Z}{%
\partial \alpha _{p}\partial \beta } \\ 
\frac{\partial ^{2}\ln Z}{\partial \alpha _{n}\partial \beta } & \frac{%
\partial ^{2}\ln Z}{\partial \alpha _{p}\partial \beta } & \frac{\partial
^{2}\ln Z}{\partial \beta ^{2}}%
\end{array}%
\right\vert .
\end{equation}%
Expressions for the evaluation of this determinant in terms of the
single-particle level densities can be found in Ref.~\cite{Moretto71}. At
some critical temperature $T^{crit}$, the gap parameter vanishes and the
excitation energy and entropy are those of a noninteracting Fermi gas, i.e.,
Eqs.~\ref{eq:A}, \ref{eq:E}, and \ref{eq:S}.

\section{DEFORMATION ENERGY}

\label{sec:deformation}

The level density will be calculated as a function of excitation energy. The
excitation energy is given in terms of the thermal contribution $E_{th}$=$%
E^{tot}(T,Q)-E^{tot}(0,Q)$ and the deformation energy $E_{def}(Q)$. In the
Strutinsky procedure\cite{Strutinsky67}, the deformation energy is given by
two terms 
\begin{equation}
E_{def}(Q)=\delta E(Q)+V_{def}(Q)
\end{equation}%
where the liquid-drop deformation energy $V_{def}(Q)$ describes the average
deformation energy with shell oscillations averaged out. The corrections $%
\delta E(Q)$ to the liquid-drop energy are determined from the
single-particle levels and have contributions from both neutrons and
protons, i.e., $\delta E=\delta E_{n}+\delta E_{p}$. Following Ref.~\cite%
{Jensen73}, we define the shell corrections as%
\begin{equation}
\delta E_{k}(Q)=E_{k}^{tot}(0,Q)-\widetilde{E_{k}^{tot}}(0,Q)
\end{equation}%
where $k$=\textit{n} or \textit{p}, and $\widetilde{E_{k}^{tot}}$ is the
total energy determined with pairing (Eq.~\ref{eq:energy_tot}), but with the
smoothed single-particle level densities 
\begin{equation}
\widetilde{g}(\varepsilon )=\int g(\varepsilon ^{\prime })F(\varepsilon
-\varepsilon ^{\prime })d\varepsilon ^{\prime }.
\end{equation}%
The smoothing function used is 
\begin{equation}
F(\varepsilon )=\frac{1}{\sqrt{\pi }\gamma }\exp \left[ -\left( \frac{%
\varepsilon }{\gamma }\right) ^{2}\right] C_{p}(\frac{\varepsilon }{\gamma })
\end{equation}%
where the smoothing range $\gamma $ must be taken to be of the order of the
intershell separation in order to washout the oscillations. The curvature
correction of order $p$=$2M$ is%
\begin{equation}
C_{p}(x)=\sum_{n=0}^{M}\frac{\left( -1\right) ^{n}}{2^{2n}n!}H_{2n}\left(
x\right) =L_{M}^{1/2}\left( x^{2}\right) .
\end{equation}%
This curvature correction\ is included to provide selfconsistancy for $%
\widetilde{g}(\varepsilon )$, i.e., a smoothed function should not be
affected by the smoothing procedure. Thus if $\widetilde{g}(\varepsilon )$
is a polynomial of order $2M+1$ or lower, it will be unchanged after the
smoothing. The functions $H_{n}$ and $L_{2p}^{1/2}$ are Hermite polynomials
and associated Laguerre polynomials, respectively.

In the original Strutinsky smoothing procedure, the smoothing parameters $%
\gamma $ and $p$ are chosen to satisfy the plateau condition\cite%
{Strutinsky67,Strutinsky69}%
\begin{equation}
\frac{d\widetilde{E_{k}^{tot}}}{d\gamma }=0,\;\frac{d\widetilde{E_{k}^{tot}}%
}{dp}=0,
\end{equation}%
over some range in both $\gamma $ and $p$. Thus in this range, the shell
correction should depend neither on the smoothing range or the order of the
curvature correction. The plateau condition can be satisfied for
single-particle levels associated with infinite potentials such as a
harmonic oscillator or an infinite square well. However for a finite-depth
potential, such as those considered in this work, the plateau condition is
often not met, i.e., one cannot find a region of $\gamma $ and $p$ over
which the shell correction is constant\cite{Brack73,Vertse98,Vertse00}.

An alternative procedure from Refs.~\cite{Vertse98,Vertse00} was tried,
however this was found to problematic in some cases. Instead the method that
is used in this work relies on the observation that the relative correction
for different deformations is independent of $\gamma $ and $p$ once $\gamma $
has a value above $\sim $1.2~$\hbar \omega $. The actual $\gamma $ value at
which the relative correction plateaus depends on the order $p$ used.
However once the plateau is reached, the relative corrections are
independent of $p$. As an example, the correction factors for neutrons
obtained with $p$=12 for various $\gamma $ values are plotted in Fig.~\ref%
{fig:correction}a. The absolute values of these corrections vary
continuously with $\gamma $ and do not plateau. However, it does have a
minimum in the interval 2.0$<\gamma <$2.5 in this example. Now apart from
the dashed curve obtained with $\gamma $=$\hbar \omega $, all other $\delta
_{n}(Q)$ curves have almost the same shape indicating the relative shell
correction is constant. To highlight this, the average value of the
correction over all deformations $\langle \delta _{n}(Q)\rangle $ is
subtracted out for each smoothing range. The remaining correction $\delta
_{n}(Q)-\langle \delta _{n}(Q)\rangle $ is plotted in the Fig.~\ref%
{fig:correction}b. All the curves for $\gamma >\hbar \omega $ now collapse
to essentially a single curve.

\begin{figure}[tbp]
\includegraphics*[scale=0.4]{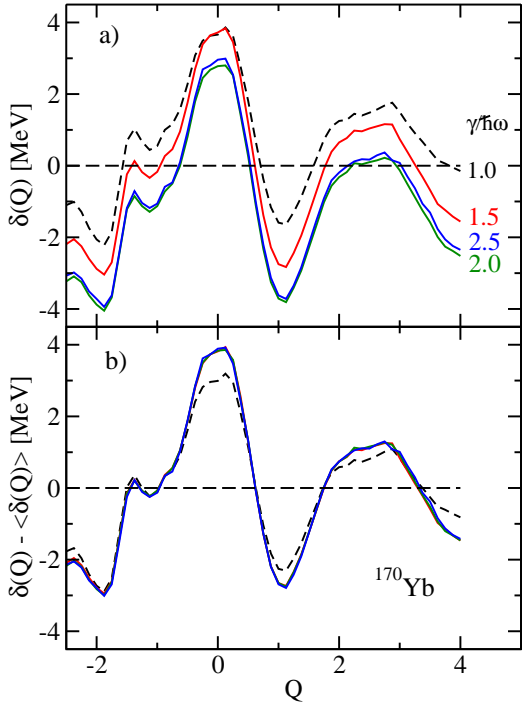}
\caption{(Color online) Variation of the neutron shell correction to the
deformation energy determined for $^{170}$Yb. a) The absolute correction
obtained with the indicated smoothing ranges $\protect\gamma $. b) The data
from a), but now the mean correction over all calculated deformations $%
\langle \protect\delta \left( Q\right) \rangle $ have been subtracted out.}
\label{fig:correction}
\end{figure}

If we cover a large enough range of deformations, the average shell
correction $\langle \delta _{k}(Q)\rangle $ is expected to be zero, thus we
have taken the values plotted in Fig.~\ref{fig:correction}b to be absolute
corrections. Thus the appropriate smoothing range is the value which causes $%
\langle \delta _{k}(Q)\rangle $=0. We expect the range of deformations
explored in this work (-2.5$<Q<$4.0) to be adequate as we always see at
least a couple of \textquotedblleft oscillations\textquotedblright\ in $%
\delta _{k}$ and thus expect the average to reflect the true average over
all deformations.

Finally the excitation energy is measured with respect to the ground-state
energy, i.e., the total excitation energy is%
\begin{equation}
E^{\ast }=E_{th}+E_{def}(Q)-\delta W.
\end{equation}%
Here the shell correction $\delta W$ represents the difference between the
liquid-drop and the minimum or ground-state deformation energy, i.e., $%
\delta W=\min \left[ E_{def}(Q)\right] $. Note that any error in the
absolute value of $\delta _{k}(Q)$ affects both $E_{def}(Q)$ and $\delta W$
equally, and therefore the excitation energy is not sensitive to the
absolute shell correction.

\begin{figure}[tbp]
\includegraphics*[scale=0.4]{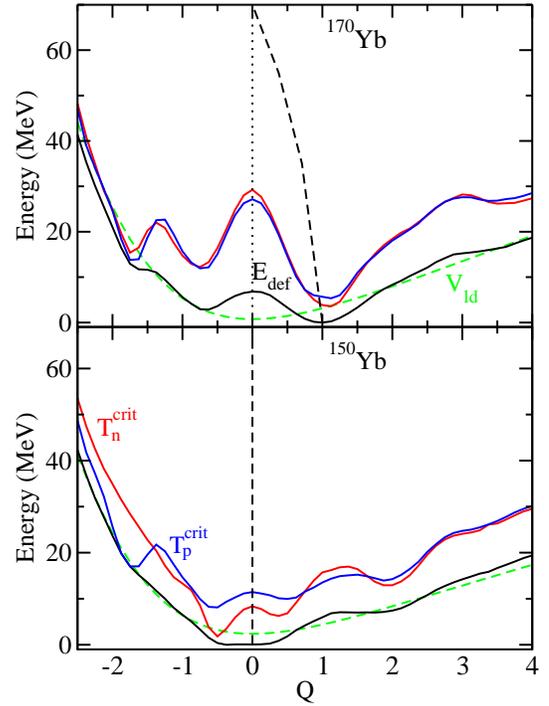}
\caption{(Color online) Variation of four calculated energies with
deformation for $^{170}$Yb and $^{150}$Yb. The curve $E_{def}$ is the
deformation energy with shell corrections. For comparison the liquid-drop
deformation energy $V_{ld}$ is also shown. The higher two curves give the
energies corresponding to the critical temperatures $T^{crit}$ for protons
and neutrons. The thick-dashed lines indicate the deformation which
maximized the level density (subtraction method) for each excitation energy.}
\label{fig:pes}
\end{figure}

The liquid-drop deformation energy is taken from Refs.~\cite{Myers66,Myers67}%
. The gap strength $G_{n,p}$ for neutrons and protons is determined from
setting $\widetilde{\Delta _{n,p}}(Q$=0,$T$=0)=$12/\sqrt{A}$~MeV\cite{Bohr75}%
. Here $\widetilde{\Delta _{n,p}}$ is the gap parameter obtained from Eq.~%
\ref{eq:gap} with the smoothed single-particle level densities $\widetilde{%
g_{n,p}}$. Examples of the deformation energy are shown in Fig.~\ref{fig:pes}
for systems with deformed ($^{170}$Yb) and spherical ($^{150}$Yb) ground
states. Also shown are the excitation energies corresponding to the critical
temperature $T_{n,p}^{crit}$ for neutrons and protons where the pairing gap
vanishes. 

\section{CALCULATIONS}

\label{sec:calculations}

\subsection{Excitation-Energy and \textit{n-p} Asymmetry Dependencies}

\label{sec:ex_dep}

The level density as function of excitation energy was calculated for
even-even nuclei with $A$=40 and 60. All such nuclei with $E_{cost}^{\min }>$%
1.9 MeV were included in the study. These include $^{40}$Ti and $^{60}$Ge
which are just beyond the proton-drip line. Calculations were also performed
for five even-even Yb nuclei from $^{150}$Yb to $^{190}$Yb covering the
range of \textit{n-p} asymmetry centered on $\beta $-stable nucleus $^{170}$%
Yb. In addition two other $A$=170 nuclei, $^{170}$Nd and $^{170}$Hg, with
extreme values of \textit{n-p} asymmetry were included. Again, $^{150}$Yb
and $^{170}$Hg are just beyond the proton drip line. Finally the heavier $%
\beta $-stable system $^{238}$U was also studied. All nuclide studied are
listed in Table~\ref{tab:list} along with their values of $\delta W$ and $%
E_{cost}^{\min }$.

\begin{table}[tbp]
\caption{Nuclei studied in this work and the value of the shell correction $%
\protect\delta W$, minimum cost $E_{cost}^{\min }$, and pairing factor $%
\protect\delta P$.}
\label{tab:list}%
\begin{ruledtabular}
\begin{tabular} {cddd}
Nucleus   & \delta W   &   E_{cost}^{\min} & \delta P \\
          & \text{(MeV)} & \text{ (MeV)} & \text{ (MeV)}     \\ \hline
$^{40}$Si &    1.1       &   1.9             &   2.7     \\
$^{40}$S  &    0.6     &     5.4           &     2.7    \\
$^{40}$Ar &    0.1     &     9.4           &     2.7    \\
$^{40}$Ca &    -0.7    &     7.4           &     2.7    \\
$^{40}$Ti &    0.6     &     3.1           &     2.6    \\
$^{60}$Ti &    -0.8    &     3.1           &     2.6    \\
$^{60}$Cr &    0.2     &     5.7           &     2.6    \\
$^{60}$Fe &    0.4     &     8.4           &     2.5    \\
$^{60}$Ni &    -0.6    &     10.1          &     2.5    \\
$^{60}$Zn &    -0.6    &     6.9           &     2.5    \\
$^{60}$Ge &    -1.6    &     3.8           &     2.6    \\
$^{170}$Nd &   0.6     &     2.5           &     2.4    \\
$^{190}$Yb &   -1.6    &     3.8           &     2.4    \\
$^{180}$Yb &   0.5     &     5.6           &     2.3    \\
$^{170}$Yb &   -0.8    &     7.6           &     2.3    \\
$^{160}$Yb &   0.7     &     9.9           &     2.2    \\
$^{150}$Yb &   -2.4    &     8.9          &     2.2    \\
$^{170}$Hg &   -2.9    &     10.0          &    2.2   \\
$^{238}$U  &   -0.7    &     5.5              &    2.3   \\
\end{tabular}
\end{ruledtabular}
\end{table}

At each deformation $Q$, the level density and excitation energy are
calculated for an array of temperatures each separated by 0.05 MeV. The
level density for a given excitation energy is then obtained from
interpolating between there results. Subsequently, the deformation of the
nucleus at each excitation energy is determined as the value which maximizes
the level density. As an example, the deformation as function of excitation
energy is plotted in Fig.~\ref{fig:pes} as the thick-dashed curves. For the
deformed ground-state system $^{170}$Yb, the deformation decreases with
excitation energy and vanishes at $E^{\ast }$=70 MeV. The spherical
ground-state system $^{150}$Yb remains spherical at all excitation energies.

\begin{figure}[tbp]
\includegraphics*[scale=0.6]{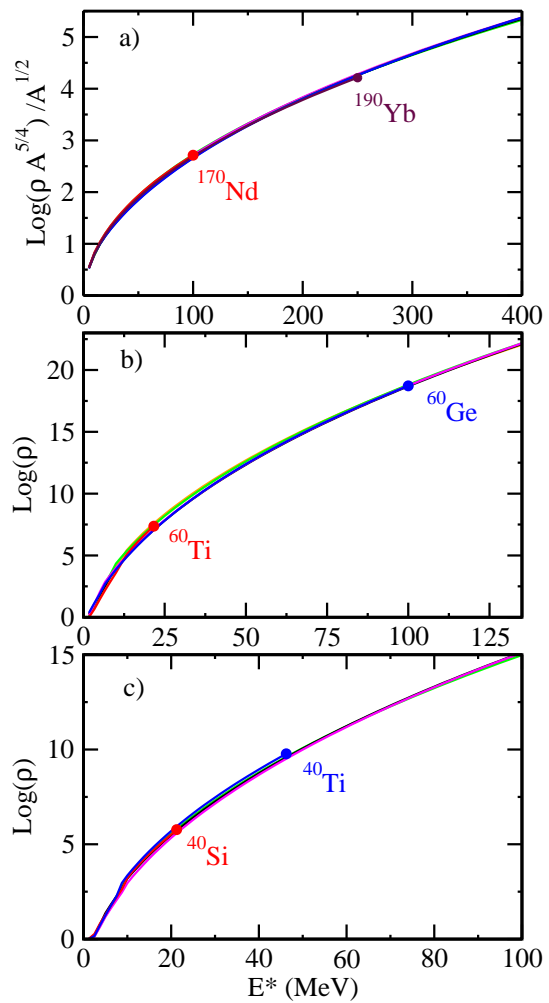}
\caption{(Color online) Variation of the nuclear level density with
excitation energy for a) the seven $A\sim $170 nuclei, b) the six $A$=60
nuclei, and c) the five $A$=40 nuclei. All are calculated up to the
excitation energy where $T$=$E_{cost}^{\min }$. These values are indicated
by the solid circular symbols if they are within the displayed ranges and
are labelled.}
\label{fig:rhosub}
\end{figure}

\begin{figure}[tbp]
\includegraphics*[scale=0.4]{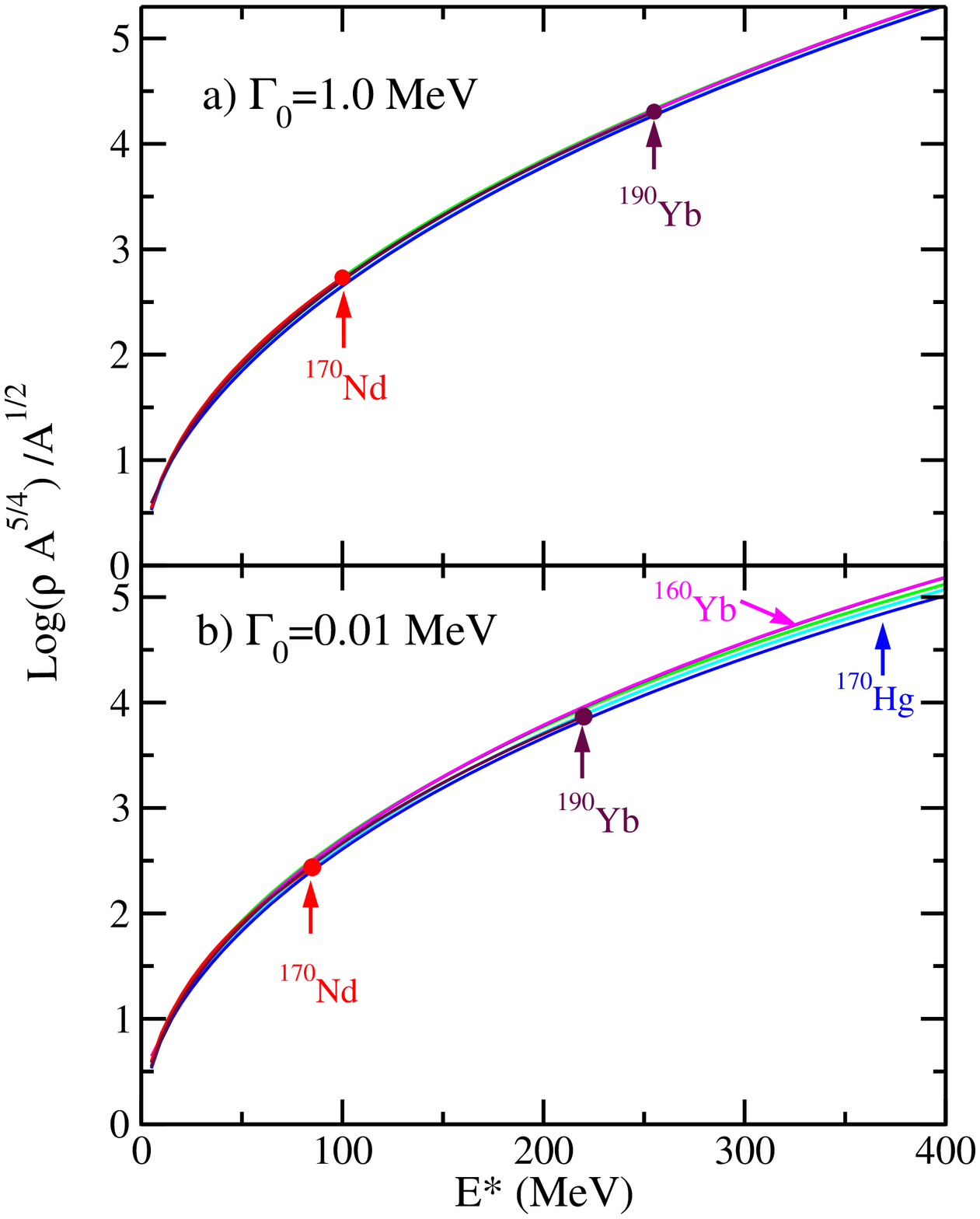}
\caption{(Color online) Same as for Fig.~\protect\ref{fig:rhosub}a, but now
the level density for $A\sim $170 is calculated with the Gamov method using
the cutoff decay widths of$\ \Gamma _{0}$=1 and 0.01~MeV. Only curves that
are distinguishable from the others are labelled.}
\label{fig:rho170}
\end{figure}

\begin{figure}[tbp]
\includegraphics*[scale=0.4]{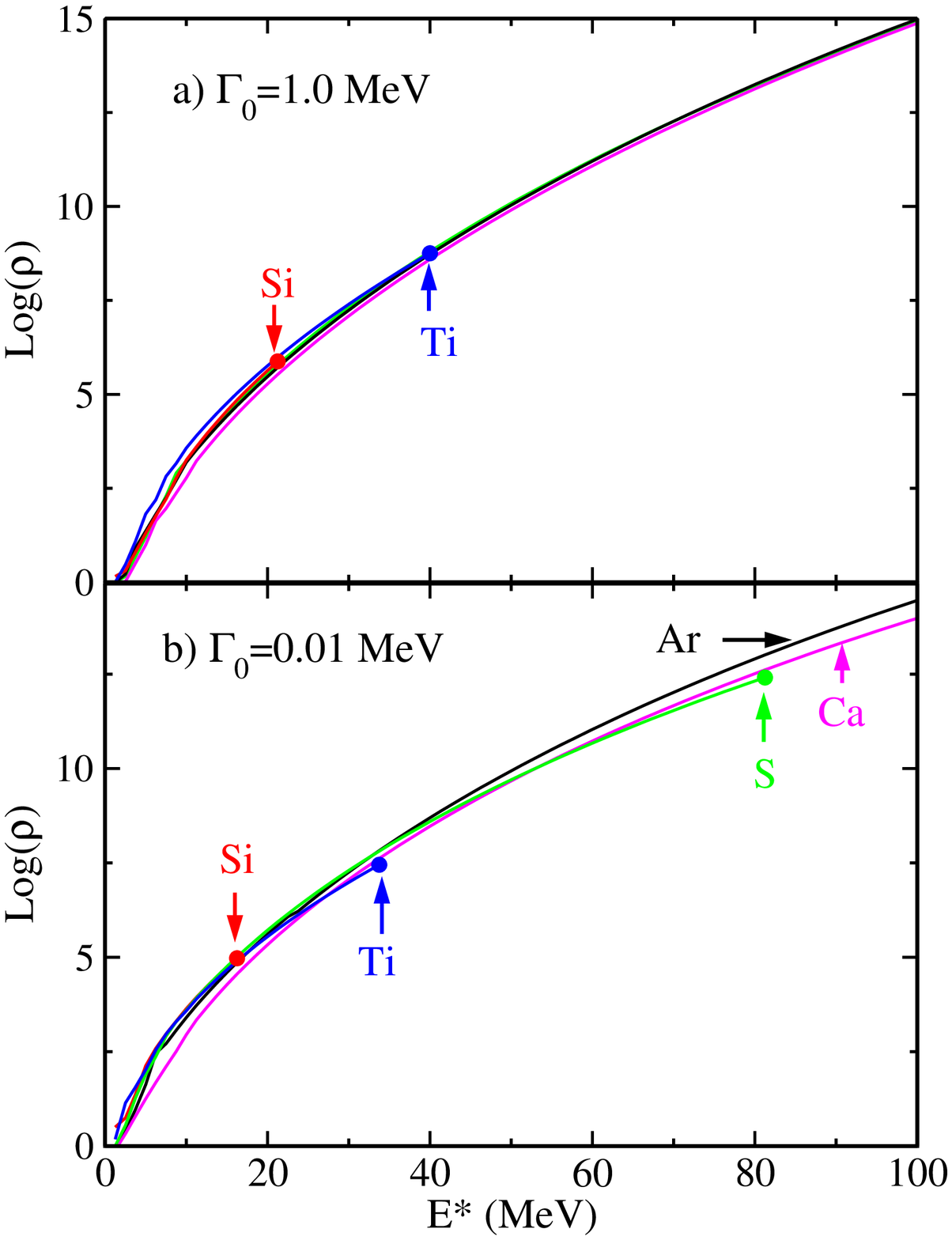}
\caption{(Color online) Same as for Fig.~\protect\ref{fig:rhosub}c, but now
the level density for $A$=40 is calculated with the Gamov method using the
cutoff decay widths of $\Gamma _{0}$=1 and 0.01~MeV. Only curves that are
distinguishable from the others are labelled.}
\label{fig:rho40}
\end{figure}

The variation of the resulting level density with excitation energy obtained
with the subtraction method is plotted in Figs.~\ref{fig:rhosub}a, b, and c
for $A\sim $170, $A$=60 and $A$=40, respectively. In Fig.~\ref{fig:rhosub}a
where $A$ is not constant, the quantity $\log \left( \rho A^{5/4}\right)
A^{1/2}$ rather than $\log \left( \rho \right) $ has been plotted to account
for the $A$ dependence based on the Fermi-gas formula with $a\propto A$.
Curves for all nuclei are only extended up to the excitation energy where $T$%
=$E_{cost}^{\min }$. It is clear from this figure, that the level density
has no substantial dependence on \textit{n-p} asymmetry, all curves with
similar $A$ values practically overlap. Similar conclusions were also
obtained with the Gamov method. For example, the level densities for $A\sim $%
170 and $A$=40 are shown in Figs.~\ref{fig:rho170} and \ref{fig:rho40},
respectively, for $\Gamma _{0}$=1.0 and $\Gamma _{0}$=0.01 MeV. Again, the
curves for similar $A$ values fall almost on top of each other. 

For $T>T_{n}^{crit}$and $T>T_{p}^{crit}$ in even-even nuclei, the excitation
energy is often backshifted by the condensation energy when comparing level
densities \cite{Moretto72}, i.e., $U$=$E^{\ast }-\delta P$ where the
smoothed condensation energy is%
\begin{equation}
\delta P=\frac{1}{2}\widetilde{\Delta _{n}}^{2}\widetilde{g_{n}}(\widetilde{%
\mu _{n}})+\frac{1}{2}\widetilde{\Delta _{p}}^{2}\widetilde{g_{p}}(%
\widetilde{\mu _{p}}).
\end{equation}%
For odd-even and odd-odd nuclei, $\delta P$ should include the pairing
correction in the semiempirical mass formula. In the comparison of level
densities in Figs.~\ref{fig:rhosub}--\ref{fig:rho40}, the role of pairing is
not important as the condensation energy is relatively constant for each
mass region (see Table~\ref{tab:list}). At high excitation energies where
shell effects are expected to be washed out, the excitation energy is also
shifted by the shell correction $\delta W$\cite{Huizenga72}. Thus at high
energies, a shifted Fermi-gas expression is often assumed. In this case, the
entropy is 
\begin{equation}
S=2\sqrt{\widetilde{a}\left( E^{\ast }-\delta P+\delta W\right) }
\label{eq:asymptotic}
\end{equation}%
where $\widetilde{a}$ is the asymptotic level-density parameter.

\begin{figure}[tbp]
\includegraphics*[scale=0.6]{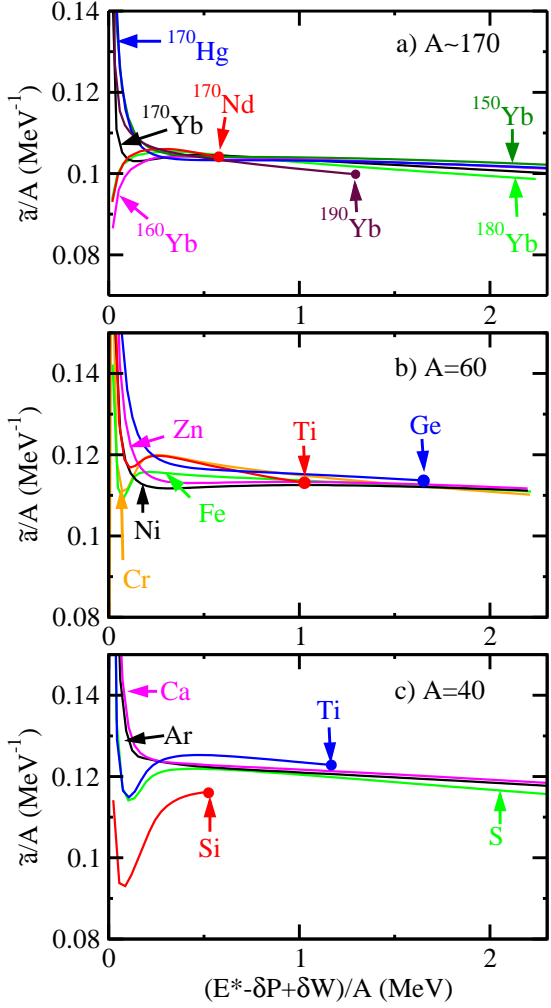}
\caption{(Color online) Level-density parameters $\widetilde{a}$ deduced
from the calculated entropy and Eq.~\protect\ref{eq:asymptotic} as a
function of the shifted excitation energy per nucleon. Results were obtained
using the subtraction method to treat the continuum. The data points
indicated on each curve are the point at which $E_{cost}^{\min }$=$T$.}
\label{fig:adensub}
\end{figure}

\begin{figure}[tbp]
\includegraphics*[scale=0.4]{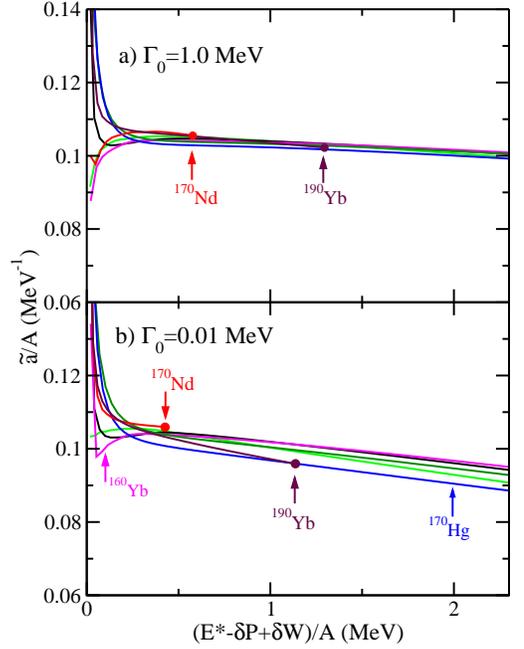}
\caption{(Color online) Same as for Fig.~\protect\ref{fig:adensub}a, but now
the results for $A\sim $170 were obtained from the Gamov method with a) $%
\Gamma _{0}$=1~MeV and b) $\Gamma _{0}$=0.01~MeV. Only curves that are
distinguishable from the others are labelled.}
\label{fig:aden170}
\end{figure}

\begin{figure}[tbp]
\includegraphics*[scale=0.4]{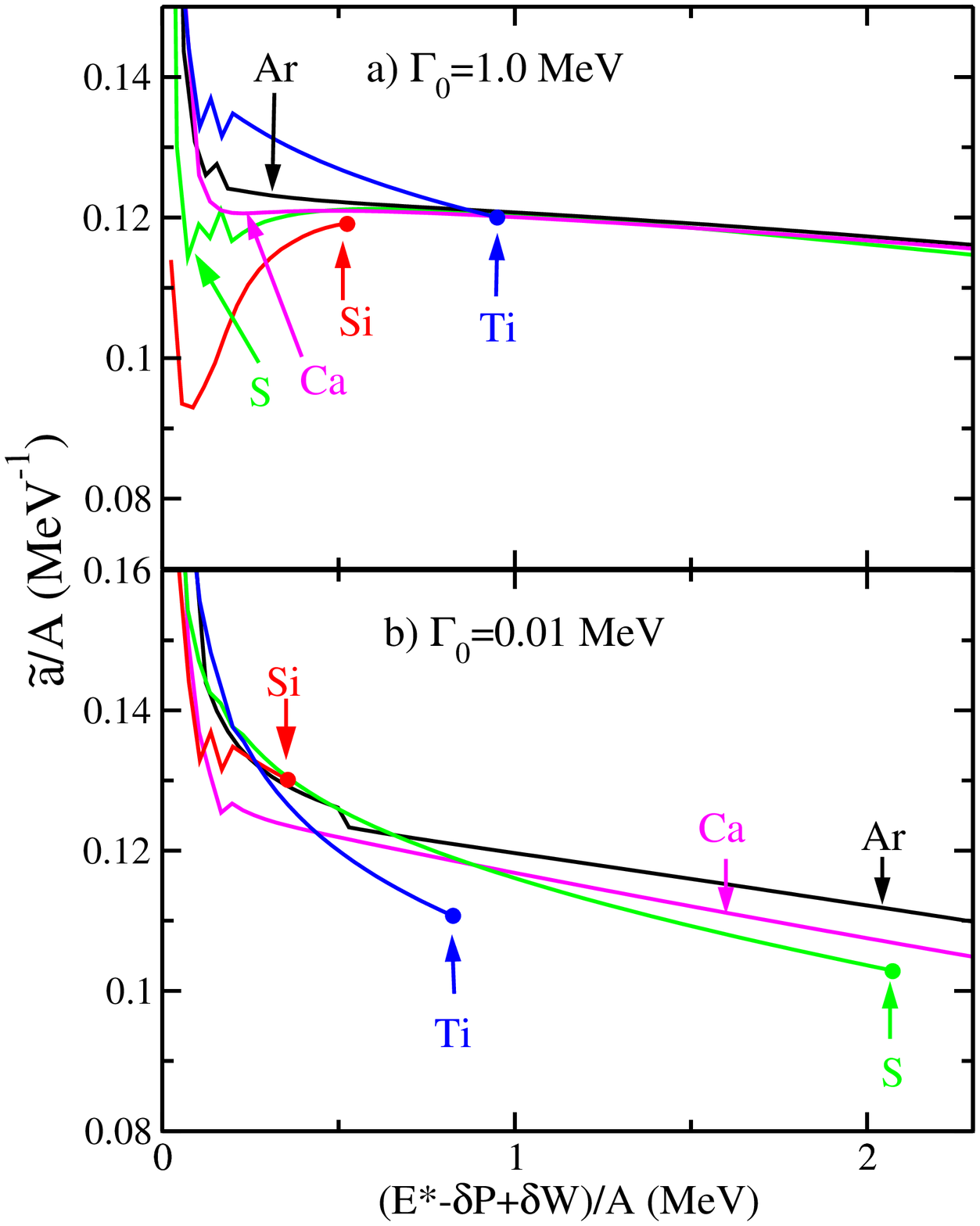}
\caption{(Color online) Same as for Fig.~\protect\ref{fig:adensub}c, but now
the results for $A$=40 were obtained from the Gamov method with a) $\Gamma
_{0}$=1~MeV and b) $\Gamma _{0}$=0.01~MeV.}
\label{fig:aden40}
\end{figure}

To see whether this formalism is consistent with the calculations of this
work, the asymptotic level-density parameter is calculated from Eq.~\ref%
{eq:asymptotic} for all excitation energies. Examples of the resulting
level-density parameters are displayed in Fig.~\ref{fig:adensub} for the
subtraction method and in Figs.~\ref{fig:aden170} and \ref{fig:aden40} for
the Gamov method. Above $\left( E^{\ast }-\delta P+\delta W\right) /A>$0.3
MeV where shell and pairing effects are expected to the quenched, the
deduced level-density parameter is rather constant. Note, Figs.~\ref%
{fig:adensub}--\ref{fig:aden40} have offset origins on the \textit{y} axis
to accentuate the difference between the different nuclei. Quite
surprisingly, the inclusion of realistic single-particle level densities
including\ continuum corrections does not cause strong deviations from the
basic Fermi-gas expression which was derived for constant $g\left(
\varepsilon \right) $. A similar conclusion was found in Hartee-Fock
calculations of $^{208}$Pb performed by Bonche, Levit, and Vautherin\cite%
{Bonche84}.


\begin{figure}[tbp]
\includegraphics*[scale=0.4]{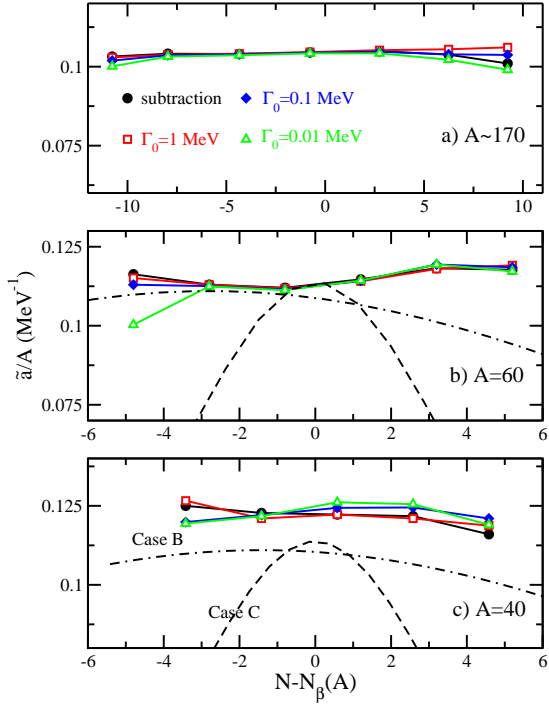}
\caption{{}(Color online) Calculated level-density parameters at $\left(
E^{\ast }-\protect\delta P+\protect\delta W\right) /A$=0.5~MeV are plotted
verses $N-N_{\protect\beta }(A)$, the neutron number separation from the $%
\protect\beta $ valley of stability . The data points were obtained with the
subtraction and Gamov methods. For the latter, the three indicated value of
the cutoff width $\Gamma _{0}$ were used. For $A$=40 and 60, the fitted
variation of the level-density parameter (cases B and C) from Ref.~%
\protect\cite{Al-Quraishi01} are shown by the dot-dashed and dashed curves,
respectively.}
\label{fig:aden}
\end{figure}

The dependence of $\widetilde{a}$ on excitation energy is not completely
flat, all calculations show some small negative slope as is expected in a
lowest-order expansion in temperature (Eq.~\ref{eq:fermiplus}). This is most
significant in the Gamov method (Figs.~\ref{fig:aden170}b and \ref%
{fig:aden40}b) with the smallest value of cutoff the width ($\Gamma _{0}$%
=0.01~MeV). The plots of the asymptotic level-density parameter also
highlight the small differences between the nuclei with the similar $A$
values which were difficult to see in Figs.~\ref{fig:rhosub}--\ref{fig:rho40}%
. The differences between the curves would become more significant if they
are extended beyond the point $T>E_{cost}^{\min }$ where the the statistical
model is problematic. For $A$=40, the asymptotic level-density parameter can
be somewhat smaller for the systems with extreme \textit{n-p} asymmetries.
For example, look at the results for $^{40}$Si in Fig.~\ref{fig:adensub}c
with the subtraction method and for $^{40}$Ti in Fig.~\ref{fig:aden40}b with 
$\Gamma _{0}$=0.01~MeV. For both of these cases, the asymptotic region is
only approach when $E_{cost}^{\min }\approx T$. To investigate the small 
\textit{n-p} asymmetry dependence more systematically, the value of $\ 
\widetilde{a}$ determined at $\left( E^{\ast }-\delta P+\delta W\right) /A$%
=0.5~MeV is plotted in Fig.~\ref{fig:aden} as a function of $N-N_{\beta
}\left( A\right) $, the distance from the $\beta $-valley of stability.
Results are shown for the subtraction method (filled circles), and the Gamov
method with three values of the maximum width; $\Gamma _{0}$=1~MeV (hollow
squares), $\Gamma _{0}$=0.1~MeV (filled diamonds), and $\Gamma _{0}$%
=0.01~MeV (hollow triangles). Generally, the deduced values of $\widetilde{a}
$ are approximately independent of which treatment of the continuum was used
and are almost constant for each mass region. In the Gamov method, the
nuclei with the extreme values of $N-N_{\beta }\left( A\right) $ show the
greatest sensitivity to $\Gamma _{0}$. In this case the values of $%
\widetilde{a}$ obtained with $\Gamma _{0}$=0.01~MeV\ are often slightly
smaller. 

\begin{figure}[tbp]
\includegraphics*[scale=0.4]{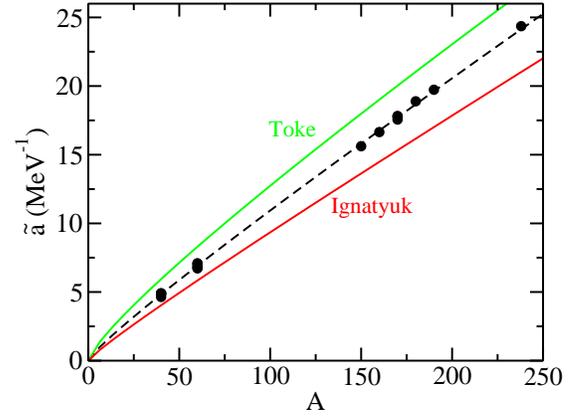}
\caption{{}(Color online) Asymptotic level-density parameter $\widetilde{a}$
as a function of nucleon number $A$. The data points are the values
determined from the calculations of this work. The dashed curve shows a fit
to these data. Curves are also shown for the prescription of T\~{o}ke and 
\'{S}wiatecki\protect\cite{Toke81} and Ignatyuk \textit{et al.}\protect\cite%
{Ignatyuk75}.}
\label{fig:littlea}
\end{figure}

The mass dependence of the level-density parameter at $\left( E^{\ast
}-\delta P+\delta W\right) /A$=0.5~MeV is displayed in Fig.~\ref{fig:littlea}
as the data points. The extracted points were fit by the commonly used
formula%
\begin{equation}
\widetilde{a}=\alpha _{v}\,A+B_{s}\,\alpha _{s}\,A^{2/3}  \label{eq:asurf}
\end{equation}%
which includes volume and surface contributions where $\alpha _{v}$ and $%
\alpha _{s}$ are the coefficients for these two quantities. The
dimensionless parameter $B_{s}$ gives the surface area of the nucleus
relative to its spherical value. As all the systems studied are spherical at 
$\left( E^{\ast }-\delta P+\delta W\right) /A$=0.5~MeV, $B_{s}$ was set to
unity. The dashed curve in Fig.~\ref{fig:littlea} shows the fit obtained
with Eq.~\ref{eq:asurf}. The fitted coefficients are $\alpha _{v}$=0.078~MeV$%
^{-1}$ and $\alpha _{s}$=0.146~MeV$^{-1}$. For comparison, curves for the
level-density parameters from T\~{o}ke and \'{S}wiatecki ($\alpha _{v}$%
=0.068~MeV$^{-1}$, $\alpha _{s}$=0.274~MeV$^{-1}$)\cite{Toke81} and Ignatyuk 
\textit{et al.} ($\alpha _{v}$=0.073~MeV$^{-1}$, $\alpha _{s}$=0.095~MeV$%
^{-1}$)\cite{Ignatyuk75} are also displayed. The fitted surface coefficient
is intermediate in value between these two other prescriptions, but closer
to Ignatyuk \textit{et al.} 

\begin{figure}[tbp]
\includegraphics*[scale=0.6]{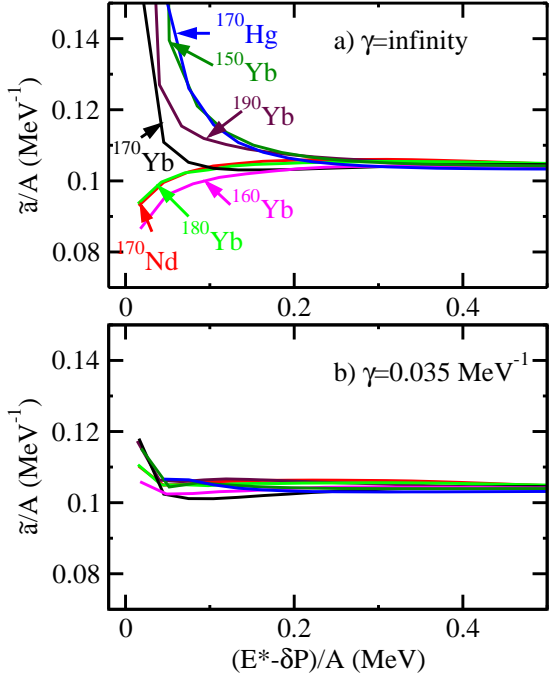}
\caption{(Color online) Asymptotic level-density parameters determined for
the $A\sim $170 systems with Eq.~\protect\ref{eq:Ignatyuk}. a) The parameter 
$\protect\gamma $ was set to infinity. These results are the same as those
shown in Fig.\protect\ref{fig:adensub}a. b) The value $\protect\gamma $%
=0.035 was obtained from minimizing the spread in the curves at low
excitation energy.}
\label{fig:adenshell}
\end{figure}

Many experimental studies have adopted the excitation-energy dependence of
the level density suggested by Ignatyuk \textit{et al}.\cite%
{Ignatyuk75,Ignatyuk79} which includes the washing out of shell effects with
increasing temperature. The entropy is expressed in terms of an
excitation-energy dependent level-density parameter, i.e., $S=2\sqrt{a(U)U}$
where 
\begin{equation}
a(U)=\widetilde{a}\left[ 1+h\left( U\right) \frac{\delta W}{U}\right] .
\label{eq:Ignatyuk}
\end{equation}%
The function $h$, determining the behavior at low excitation energies, is
given by $h(U)$=1-$\exp \left( -\gamma U\right) $. The parameter $\gamma $
gives the energy scale over which shell effects are washed out. At high
excitation energies where $h\rightarrow 1,$ the Ignatyuk formalism leads to
the expected dependence of Eq.~\ref{eq:asymptotic}. To determine how well
Eq.~\ref{eq:Ignatyuk} can describe the entropy calculated in this work,
asymptotic level-density parameters were determined at each excitation
energy from Eq.~\ref{eq:Ignatyuk} and the parameter $\gamma $ was adjusted
to minimize the spread of these deduced $\widetilde{a}$ values at low
excitation energies. For the $A\sim $170 nuclei, $\gamma $ was determined by
this procedure as 0.035 MeV$^{-1}$ and the $\widetilde{a}$ values are
plotted in Fig.~\ref{fig:adenshell}b. These are to be compared to the $%
\widetilde{a}$ values obtained for $\gamma $=$\infty $ in Fig.~\ref%
{fig:adenshell}a. The condition $\gamma $=$\infty $ corresponds to $h$=0 and
the latter values are the same as deduced from Eq.~\ref{eq:asymptotic} and
plotted in Fig.~\ref{fig:adensub}a. The spread in the $\widetilde{a}$ values
at low excitation energies observed in Fig.~\ref{fig:adenshell}a are almost
removed in Fig.~\ref{fig:adenshell}b and thus this indicates that the
Ignatyuk formalism describes the fading out of shell effects adequately for
this mass region. The deduced value of $\gamma $ is of similar magnitude to
the value 0.05 obtained by Ignatyuk \textit{et al.}\cite{Ignatyuk75} by
fitting neutron resonance data. Schmidt \textit{et al.}\cite{Schmidt82} have
extracted a mass-dependent value of $\gamma $ and, for $A$=170, they find $%
\gamma $=0.045. Again of similar magnitude to the value of this work. 

For the lighter mass regions ($A$=40 and 60), a similar reduction in the
spread of the deduced $\widetilde{a}$ values was not achieved. Thus for
these light systems, the description of the level density in the region
where shell effects are still important is more complex than this Ignatyuk
treatment.

\subsection{Deformation Dependence}

\begin{figure}[tbp]
\includegraphics*[scale=0.4]{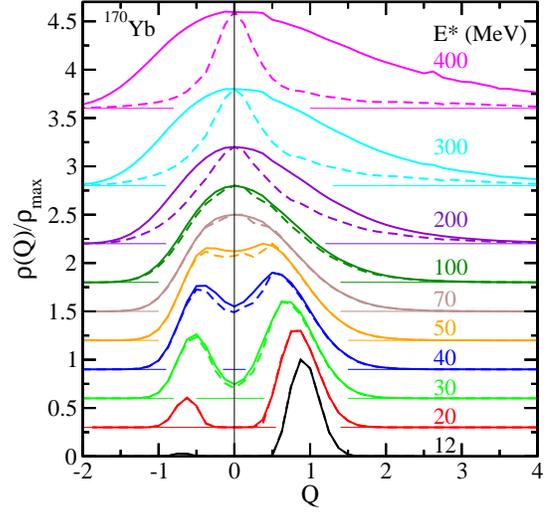}
\caption{(Color online) Deformation dependence of the level density
calculated for $^{170}$Yb at the indicated excitation energies. The level
densities are normalized to the maximum $\protect\rho_{max}$ for that
excitation energy. For clarity the results for each successive excitation
energy are shifted up along the \textit{y} axis. The thin lines in each case
correspond to the shifted \textit{x} axis. Results obtained with the
subtraction method are shown as the thick-solid curves, while the dashed
curve were obtained from the Gamov method with $\Gamma _{0}$=1~MeV.}
\label{fig:def170}
\end{figure}

The level density of strongly deformed nuclei is of interest. In the
statistical model, the fission decay rate is determined from the level
density of the deformed saddle-point configuration. The deformation
dependence of the level density is also needed to determine the equilibrium
shape distribution of compound nuclei\cite{Moretto71}. This distribution can
be important in determining the emission rates of $\alpha $ and heavier
fragments\cite{Charity00}. The deformation dependence for $^{170}$Yb at
various excitation energies is displayed in Fig.~\ref{fig:def170}. To
highlight the deformation dependence, the level densities are normalized to
the maximum value for that excitation energy. The results obtained with the
subtraction method are shown as the solid curves, while the dashed curves
are from the calculation with the Gamov method ($\Gamma _{0}$=1~MeV). At the
lowest excitation energies, the level density is largest for deformations
close to the ground-state value ($Q$=0.875 for $^{170}$Yb). For this nucleus
at these excitation energies, the continuum is not sampled significantly and
the results for the two methods are almost identical. At an intermediate
energy ($\sim $100 MeV), shell effects have melted and the level density
peaks for spherical shapes but the distribution is quite broad. Again, the
results are similar for the two methods. However at higher excitation
energies where the continuum becomes more important, the results obtained
with the two methods are quite different. For the subtraction method (solid
curves), the dependence on deformation near sphericity decreases. The curves
become broader with increasing excitation energy as expected when the
temperature increases. Contrary to this, the dashed curves obtained with the
Gamov method becomes narrower. The underlying reason for this behavior can
be traced to the variation of the resonance widths with deformation
displayed in Fig.~\ref{fig:width} and discussed in Sec.~\ref{sec:results}.
As a spherical system is deformed, then on average, the widths of the
resonances increase leading to a decrease in the single-particle level
density in the Gamov method. Thus, this behavior leads to a favoring of
spherical shapes. 

\begin{figure}[tbp]
\includegraphics*[scale=0.4]{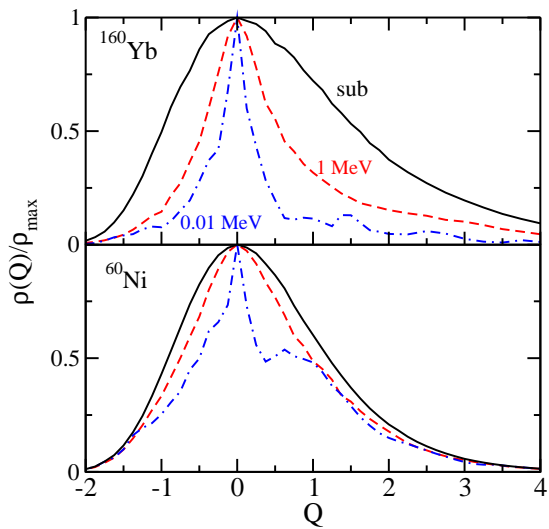}
\caption{(Color online) Deformation dependence of the normalized level
density as in Fig.~\protect\ref{fig:def170}. Results are shown for $^{160}$%
Yb ($E^{\ast }$=300~MeV) and $^{60}$Ni ($E^{\ast }$=112~MeV) with the
subtraction method (solid curves) and with the Gamov method for $\Gamma _{0}$%
=1~MeV (dashed curves) and 0.01 MeV (dot-dashed curves).}
\label{fig:defcom}
\end{figure}

The strong peaking of the level density at sphericity for the Gamov method
is quite general. Results are shown in Fig.~\ref{fig:defcom} for $^{160}$Yb
and $^{60}$Ni at $E^{\ast }/A$ =1.875 MeV. In both cases, the continuum
contributions are significant. Curves are shown for the subtraction method
(solid) and for $\Gamma _{0}$=1 MeV(dashed) and $\Gamma _{0}$=0.01~MeV
(dot-dashed). For both nuclei, the level density distributions obtained with
the Gamov method are narrower than those from the subtraction method.
However, the peaking at sphericity is even stronger for smaller values of
the cutoff width $\Gamma _{0}$. Also, the effect is stronger for the heavier
system. These dependencies are quite general. 

\begin{figure}[tbp]
\includegraphics*[scale=0.4]{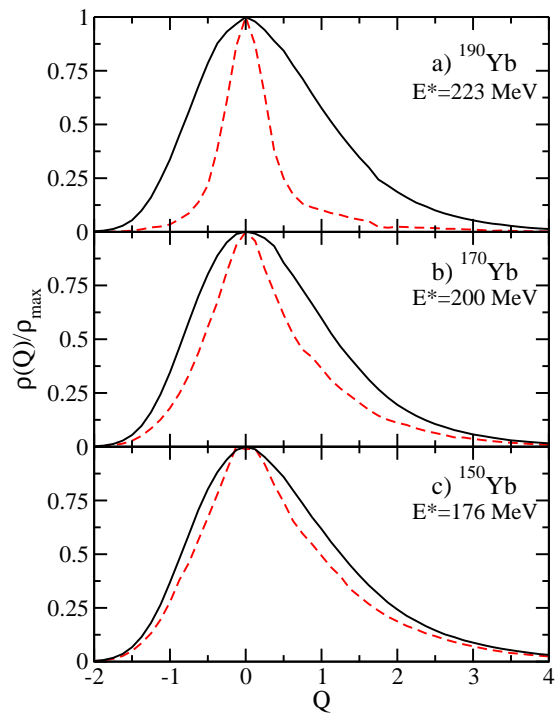}
\caption{(Color online) Deformation dependence of the normalized level
density for three Yb isotopes at the indicated excitation energies. Solid
and dashed curves were obtained with the subtraction and Gamov ($\Gamma _{0}$%
=1~MeV) methods, respectively.}
\label{fig:defa}
\end{figure}

The degree to which the spherical shape is favored in the Gamov method also
depends on the \textit{n-p} asymmetry. Figure~\ref{fig:defa} displays the
deformation dependence of the level density for three Yb isotopes all at $%
E^{\ast }/A$=1.18~MeV. The dashed curves, calculated with the Gamov method,
indicate that the effect is stronger for the very neutron-rich $^{190}$Yb
isotope. Because of the small neutron separation energy for this system, the
importance of the positive-energy neutron levels is much greater than for
the $\beta $-stable $^{170}$Yb system. The proton-rich $^{150}$Yb system
shows an even smaller effects than the $\beta $-stable nucleus. The most
important positive-energy proton levels are below Coulomb the barrier and
thus they are all narrow compared to the cutoff width $\Gamma _{0}$. As a
general rule, modifications induced by deformation are of lesser importance
for protons compared to neutrons. Of course for very light nuclei, the
Coulomb barrier is smaller and its ability to suppress these deformation
effects is reduced. 

The deformation dependence from the Gamov method has important consequences
for fission. The fission decay width is determined from the ratio of the
level densities at the saddle-point and equilibrium configurations. The
favoring of spherical nuclei at high excitation energies will lead to a
suppression of the fission width and thus an increase in the probability of
the competing evaporative decay modes. The total fission cross section will
therefore be reduced and, for events that do fission, it will occur later in
the decay cascade. Such effects have been observed experimentally.
Measurements of pre and postscission multiplicities of evaporated particles
in coincidence with fission fragments have shown that these fragments are
created with little excitation energy even when the initial CN excitation
energy is large\cite{Hilscher92}. The standard interpretation of these
results is in term of dynamical effects\cite{Hilscher92}, but it is clear
that if one adopts the Gamov method, then it can explain part, or possibly
most, of the experimental observations. The magnitude of the predicted
effect will depend on the value of $\Gamma _{0}$.

\begin{figure}[tbp]
\includegraphics*[scale=0.4]{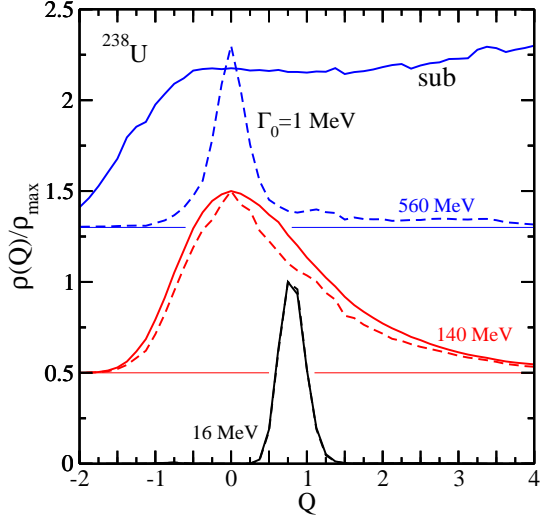}
\caption{(Color online) Deformation dependence of the normalized level
density for $^{238}$U as in Fig.~\protect\ref{fig:def170}.}
\label{fig:def238}
\end{figure}

Fission is most important for the heavier systems and thus it is of interest
to examine the deformation dependence determined for $^{238}$U. This in
shown in Fig.~\ref{fig:def238} for three excitation energies. For the lowest
value ($E^{\ast }$=16~MeV), the level density is again largest for
deformations around the ground-state value. There is again no difference
between the subtraction (solid curve) and Gamov (dashed curve) methods. (The
curves are indistinguishable). At $E^{\ast }$=140~MeV shell effects have
melted and, in both methods, the level density is largest for spherical
systems. Only a small difference between the two methods is observed. Again
at the highest excitation energy ($E^{\ast }$=560 MeV), the two methods give
very difference results. The Gamov method is strongly peaked at sphericity.
In contrast now, the results with the subtraction method show this nucleus
is unstable with respect to prolate deformations, i.e., the level density
increases with increasing $Q$. This is basically a fission instability,
however to fully treat fission one should include more shape degrees of
freedom. If the level-density parameter is deformation dependent as in Eq.~%
\ref{eq:asurf}, then the fission barrier is temperature dependent\cite%
{Charity96}. The result obtained with the subtraction method therefore
represents the situation where the temperature-dependent fission barrier has
vanished. In terms of level density, there is no saddle-point, i.e., a
configuration of low level density which represents a bottleneck through
which the system must pass in order to fission. Therefore, fission stability
can be quite different for the two methods of treating the continuum. 

\begin{figure}[tbp]
\includegraphics*[scale=0.4]{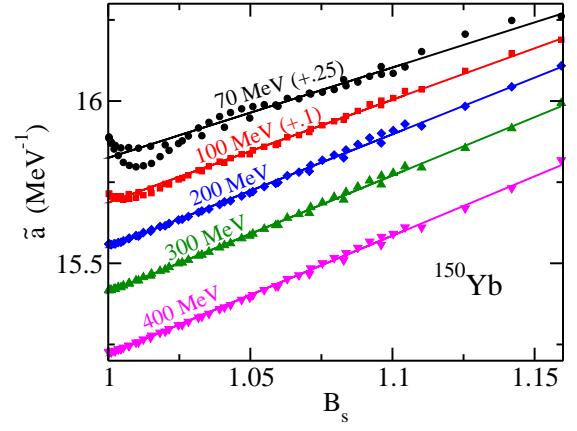}
\caption{(Color online) Asymmptotic level-density paramters deduced for
deformed $^{150}$Yb nuclei (oblate and prolate shapes) as a function of the
relative surface area $B_{s}$. Results are shown for the indicated thermal
excitation energies. For clarity, results at some excitation energies have
been shifted up along the \textit{y} axis by the indicated amounts.}
\label{fig:def}
\end{figure}

\begin{figure}[tbp]
\includegraphics*[scale=0.4]{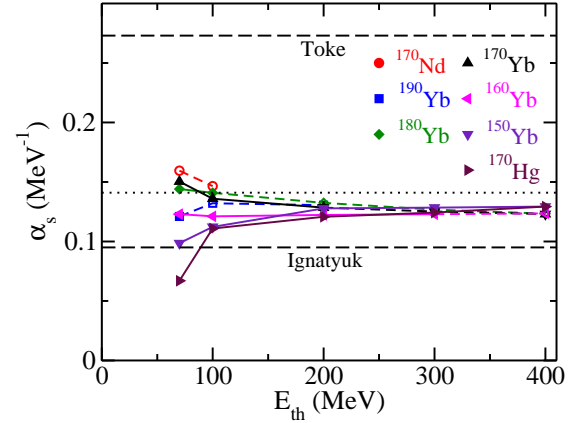}
\caption{(Color online) Surface coefficient of the level-density parameter
obtained for the $A\sim $170 systems as a function of excitation energy. The
dotted line indicates the value obtained from the fit in Fig.~\protect\ref%
{fig:littlea}. Other values of the surface coefficient, from the
prescriptions of T\~{o}ke and \'{S}wiatecki\protect\cite{Toke81} and
Ignatyuk \textit{et al.}\protect\cite{Ignatyuk75}, are also indicated.}
\label{fig:asurf}
\end{figure}

As the deformation-dependence of the level-density parameter plays an
important role in fission, the applicability of Eq.~\ref{eq:asurf} was
investigated for the subtraction method. (It is clearly not applicable for
the Gamov method). For a given excitation energy and deformation, the
asymptotic level-density parameter was determined as%
\begin{equation}
S=2\sqrt{\widetilde{a}\left( E^{\ast }-\delta P+\delta W-V_{def}(Q)\right) }
\end{equation}%
assuming shell and pairing effects are washed out. This differs from Eq.~\ref%
{eq:asymptotic} in that now the liquid-drop deformation energy has also been
subtracted from the excitation energy. Results obtained for $^{150}$Yb are
displayed in Fig.~\ref{fig:def} as the data points for various values of $%
E_{th}=E^{\ast }-\delta P+\delta W-V_{def}(Q)$, the asymptotic thermal
excitation energy. The extracted values are plotted against $B_{s}$, the
relative surface area ($B_{s}$=1 is sphericity). Except for the lowest value
of $E_{th}$ where shell oscillations are still present, they increase almost
linearly with $B_{s}$. The solid curves display linear fits to the extracted
values and, from the fitted slopes, the surface coefficient $\alpha _{s}$
(Eq.~\ref{eq:asurf}) can be determined. Ignoring $E_{th}$=70~MeV, the
slopes, and thus the $\alpha _{s}$ coefficients, are almost independent of
excitation energy. The $\alpha _{s}$ coefficients obtained from all
calculated $A\sim $170 nuclei are plotted in Fig.~\ref{fig:asurf}. Apart
from the lowest excitation energies where shell effects are still important,
all the $\alpha _{s}$ coefficients are similar, almost independent of
excitation energy and \textit{n-p} asymmetry. The values of $\alpha _{s}$
determined by this procedure are quite similar to the value obtained from
fitting the $A$ dependence of $\widetilde{a}$ in Sec.~\ref{sec:ex_dep} .
This value is indicated by the dotted line in Fig.~\ref{fig:asurf}. Clearly
Eq.~\ref{eq:asurf} provides a reasonably consistent description of the $A$
and deformation dependencies of the level-density parameter for the
subtraction method. The small difference between the extracted values of $%
\alpha _{s}$ from the two procedures may be due to the fact that curvature
and higher-order corrections to the level-density parameter\cite{Toke81}
have been ignored. Thus, these must be small (at least for the deformations
considered) in order to get such good agreement from the two procedures.


One final note, $\alpha _{s}$ depends very much on details of how the
mean-field potential changes with deformation. In Sec.~\ref{sec::cctheory},
the diffuseness parameter $d$ of the nuclear potential was made angle
dependent (Eq.~\ref{eq:diffuseness}) so that the diffuseness perpendicular
to the nuclear surface is constant. The parameter $d$ is actually the
diffuseness along the radial direction. If instead $d$ is set to be constant
as in many studies, then the extracted $\alpha _{s}$ values were found to be
negative! In this case, the mean diffuseness perpendicular to the nuclear
surface decreases with deformation. As the level-density parameter is quite
sensitive to the diffuseness, this leads to the calculated reduction of the
level-density parameter with deformation. Thus for large deformations, it is
important to make $d$ angle dependent.

\section{DISCUSSION}

\label{sec:discussion}

In the preceding section, it was shown that the two methods for calculating
the contribution from the positive-energy single-particles levels give
similar results, apart from the deformation dependence. The difference
between the two methods are the inclusion of both the negative background
and the wide resonances in the subtraction method. These two contributions
have opposite effects on the level density, and thus it seems they partially
cancel each other. Also for heavy systems with large Coulomb barriers, the
low, positive-energy proton states are all narrow resonances and thus give
almost identical results for the two methods. Thus the proton-rich side of
the chart of nuclides is less sensitive than the neutron-rich side to the
continuum, at least for the heavier systems.

If the Gamov method is considered preferable, then consideration must be
given to the value of $\Gamma _{0}$. Possibly $\Gamma _{0}$ is related to
the CN\ lifetime, i.e., only Gamov levels of lifetime greater than the CN
should be considered. As the CN lifetime decreases with excitation energy,
then $\Gamma _{0}$ should increase. An increasing value of$\ \Gamma _{0}$
with excitation energy would lead to a reduction in the predicted strong
peaking of the level density at sphericity. Another possibility is that $%
\Gamma _{0}$ should be related to the average spreading width of levels near
the Fermi surface. This would lead to a much smaller excitation-energy
dependence of $\Gamma _{0}$.

It is important to remember that the results of this work relate to how the
treatment of positive-energy single-particle levels affects the level
density in the independent-particle model. Apart from pairing, the
calculations did not include any other many-body effects. Also they did not
allow for selfconsistancy between the assumed nuclear potential and
predicted density distributions of the nucleons. These effects may to lead
to deviations from the predicted behavior. In fact, experimentally it is
known that the level density for hot Yb nuclei cannot be described by a
backshifted Fermi-gas expression, but an important excitation-energy
dependence of the level-density parameter is needed\cite{Charity03}. The
latter is consistent with the variation of the frequency-dependent effective
nucleon mass with temperature\cite{Bortignon87}.

Some of the effects ignored in this work may influence the \textit{n-p}
asymmetry dependence. For example, if the diffuseness parameter of the
nuclear potential increases for nuclei close to the drip lines, then this
will enhance the level-density parameter for these systems. Also differences
between the neutron and proton effective masses and their dependence on
asymmetry may also be important.

Al-Quraishi \textit{et al.}\cite{Al-Quraishi01} had suggested that a
restriction of the positive-energy states to narrow resonances, as in the
Gamov method, would lead to an important \textit{n-p} asymmetry dependence
of the level-density parameter. They fit the density of known levels for 20$%
\leq A\leq $70 with the form%
\begin{equation}
a_{C}=a_{1}A\exp \left\{ a_{2}\left[ Z-Z_{\beta }\left( A\right) \right]
^{2}\right\}  \label{eq:caseC}
\end{equation}%
where $Z_{\beta }\left( A\right) $ is the proton number of the $\beta $%
-stable nucleus of nucleon number $A$ and $a_{1}$ and $a_{2}$ are the fit
parameters. This is called case C in Ref.~\cite{Al-Quraishi01}. The dashed
curves in Figs.~\ref{fig:aden}b and \ref{fig:aden}c show $a_{C}$ for $A$=60
and 40, respectively. If this strong \textit{n-p} asymmetry dependence is
real, the calculations of this work suggest it cannot be explained by the
treatment of the continuum as assumed in the justification of Eq.~\ref%
{eq:caseC}. Al-Quraishi \textit{et al. }also considered another fit (case B)
based on isospin considerations:%
\begin{equation}
a_{B}=a_{3}A\exp \left[ a_{4}\left( N-Z\right) ^{2}\right]  \label{eq:caseB}
\end{equation}%
where now $a_{3}$ and $a_{4}$ are the fit parameters. The dot-dashed curves
in Figs.~\ref{fig:aden}b and \ref{fig:aden}c show the resulting values of $%
a_{B}$. In this case, the \textit{n-p} asymmetry dependence is not as strong
as in case C, but still stronger than our calculations for $A$=60. The large
difference between cases B and C suggests the fits do not constrain the
level-density parameter for very neutron and proton-rich nuclei. In Ref.~%
\cite{Charity03}, no significant asymmetry dependence of the level-density
parameter was observed for $^{152}$Yb and $^{160}$Yb CN with excitation
energy greater than 100~MeV. This is consistent with the dependence
calculated with both methods of this work (see Figs.~\ref{fig:aden}a).

It is also important to note again that the small \textit{n-p} asymmetry
dependence observed in these calculations is only true for $T<E_{cost}^{\min
}$. Extending the calculations above $T=E_{cost}^{\min }$ leads to a much
larger dependence, although this region is not relevant to the statistical
model. In Ref.~\cite{Mustafa92} is was noted that for calculations where the
number of single-particle levels is finite, such as in the Gamov method,
then there is a maximum excitation energy of the CN. The level density as a
function of excitation energy must peak and then approach zero at this
maximum excitation energy. The peak value of the level density corresponds
to $T$=$\infty $ and higher excitation energies have negative temperatures.
The details of this behavior would be very dependent on the \textit{n-p}
asymmetry. Although a negative temperature may be appropriate for the CN, it
is certainly not meaningful for the gas. Thus our model of CN in equilibrium
with the surrounding gas breaks down. However, it is not clear that this is
of any relevance for the statistical model.

\section{CONCLUSIONS}

\label{sec:conclusion}

The effects of continuum positive-energy neutron and proton levels on the
nuclear level density has been investigated. The use of the
independent-particle model allowed for a broad survey of how these continuum
corrections modify the level density over the entire chart of nuclides. Two
methods for calculating the contributions of these positive-energy levels
were investigated. In the subtraction method, the single-particle level
density is determined from the scattering phase shifts. The resulting
single-particle level density has contributions from narrow and wide
resonances and a negative background. In the Gamov method, the
single-particle level density is calculated from the Gamov states and only
the contributions from the narrow resonances are considered. From the bound
states and these two prescriptions for the positive-energy states, the
entropy and level density are calculated as a function of temperature and
excitation energy. These calculations ignored all many-body effects apart
from the pairing interaction. At large excitation energies where shell
effects melt, the level density followed a backshifted Fermi-gas expression.
Also, the deduced level-density parameters were quite similar for the two
methods. They depend on $A\ $with very little dependence on the \textit{n-p}
asymmetry of the nucleus. The biggest asymmetry dependence was for the very
exotic systems near the drip lines where a small reduction in the
level-density parameter was sometimes found. For the heavier systems, the
prescription of Ignatyuk \textit{et} \textit{al.}\cite{Ignatyuk75} accounted
for the variation in level density at low excitation energies where shell
effects are still important.

The largest differences arising from the use of the two methods was the
predicted deformation dependence of the level density. At high excitation
energies, the Gamov method predicted the level density peaked strongly for
spherical systems whereas in the subtraction method the deformation
dependence was rather flat near sphericity. This suppression in the relative
level densities of deformed to spherical systems in the Gamov method would
lead to a reduction in the predicted fission width and may help explain the
large prescission light-particle multiplicities observed in fission reactions%
\cite{Hilscher92}.

\begin{acknowledgments}
We wish to acknowledge many informative discussions with Prof. Ron Lovett
and Prof. Willem Dickhoff. This work was supported by the Director, Office
of High Energy and Nuclear Physics, Nuclear Physics Division of the U.S.
Department of Energy under contract number DE-FG02-87ER-40316.
\end{acknowledgments}

\end{document}